\documentclass[11pt,]{article}
\usepackage{authblk}
\usepackage{fullpage}
\usepackage{amssymb,amsmath}
\usepackage[utf8x]{inputenc}
\usepackage{color,soul}
\usepackage[T1]{fontenc}
\usepackage{siunitx}
\usepackage[version=3]{mhchem}
\usepackage{xr}
\usepackage{mathtools}
\usepackage{bm}
\usepackage{pdfpages}
\usepackage{graphicx}
\usepackage{dcolumn}
\usepackage{bm}
\usepackage{color}
\usepackage{tikz}
\usepackage{setspace}
\usepackage[title]{appendix}

\usepackage{float}
\usepackage{mathtools}
\usepackage{braket}
\usepackage[center,font=small,labelfont=bf]{caption}

\usepackage[square,numbers,sort&compress]{natbib}
\usepackage[left]{lineno}

\usepackage{setspace}
\usepackage[unicode=true]{hyperref}
\hypersetup{colorlinks=true,
            citecolor=blue,
            linkcolor=blue}
\numberwithin{equation}{section}

\usepackage{graphicx}
\graphicspath{{figures/}}
\makeatletter
\def\ScaleWidthIfNeeded{%
 \ifdim\Gin@nat@width>\linewidth
    \linewidth
  \else
    \Gin@nat@width
  \fi
}
\def\ScaleHeightIfNeeded{%
  \ifdim\Gin@nat@height>0.9\textheight
    0.9\textheight
  \else
    \Gin@nat@width
  \fi
}
\makeatother
\setkeys{Gin}{width=\ScaleWidthIfNeeded,height=\ScaleHeightIfNeeded,keepaspectratio}%

\begin{document}

\title{{Flow spatial structure determines pattern instabilities in nonlocal models of population dynamics.}}

\author[1,2,3]{Nathan O. Silvano}
\author[4,5]{João Valeriano}
\author[3]{Emilio Hernández-García}
\author[3]{Cristóbal López}
\author[1,5]{Ricardo Martinez-Garcia\footnote{Corresponding author: r.martinez-garcia@hzdr.de}}

\affil[1]{\footnotesize Center for Advanced Systems Understanding (CASUS) -- Helmholtz-Zentrum Dresden-Rossendorf (HZDR), Görlitz, Germany}
\affil[2]{Departmento de Física Teórica, Universidade do Estado do Rio de Janeiro, Rio de Janeiro, RJ, Brazil.}
\affil[3]{Institute for Cross-Disciplinary Physics and Complex Systems (IFISC), CSIC-UIB, Palma de Mallorca, Spain}
\affil[4]{Turing Center for Living Systems, Aix-Marseille University, CNRS, CINAM, Marseille, France}
\affil[5]{ICTP South American Institute for Fundamental Research \& Instituto de F\'isica Te\'orica, Universidade Estadual Paulista - UNESP, São Paulo SP, Brazil}

 \date{\vspace{-10ex}}

\maketitle

\normalsize
\begin{abstract}
{We investigate how environmental flows influence spatial pattern formation and population dynamics using two nonlocal models of population dynamics, which we couple to two different stationary flows. Combining numerical simulations and analytical approximations, we show that the spatial structure of the flow’s velocity field determines the pattern formation instability. For a simple shear flow, where one of the primary axes of the population pattern can become aligned with the flow, the onset of pattern formation remains unaffected. In contrast, a vortex flow delays the pattern instability relative to the no-flow case. The velocity field, therefore, interacts with the spatial feedbacks responsible for pattern formation in complex ways, which also leads to different oscillatory time series of population abundance. In some cases, the population undergoes regular oscillations with a characteristic frequency, while in others, the dynamics exhibits long erratic transients with no well-defined period before settling into a more regular behavior.}
\end{abstract}

\section{Introduction}\label{sec1}
 Populations very often form self-organized spatial patterns caused by nonlinear interactions among organisms and between organisms and their environment \citep{Martinez-Garcia2022}. Examples of these patterns appear on a broad range of observational scales, from cell populations to landscapes \citep{Pringle2017, Karig2018, Rietkerk2021, Martinez2023}, and both in marine and terrestrial environments \citep{Rietkerk2008, Zhong2012, Ruiz2017}. Across this variety of systems, patterns often impact several ecological processes such as population growth and extinction \citep{Sun2020, Surendran2020, Jorge2024}, species interactions \citep{DeRoos1998, Maciel2021, Eigentler2021, MartinezCalvo2023}, or evolutionary dynamics \citep{Wakano2009, Park2019}. A lot of research has therefore focused on understanding, mainly using theoretical models, how different feedbacks between processes controlling population growth and movement affect pattern formation and their ecological and evolutionary consequences \citep{Rietkerk2008,Martinez-Garcia2022}. 

Most of these studies assume that organisms move randomly, either with constant or density-dependent diffusion \citep{Mimura1981,MartinezGarcia2015,Liu2014,Liu2016}, or exhibiting some sort of active movement \citep{Farrell2012,Cates2015}. Other spatial processes affecting population mixing, such as those induced by environmental flows in aqueous environments, also have important ecological and evolutionary consequences. For example, planktonic microorganisms live in environments where rapidly changing flow velocities can create different levels of population mixing \citep{Young2001,Guasto2012,Neufeld2009} and impact encounter rates \citep{visser2006}. This dependence of the encounter rates on flow properties has important consequences for population growth \citep{Ser-Giacomi2023, Comesana2021} and patchiness \citep{Martin2003}, determines the outcome of species interactions \citep{Pigolotti2012, Uppal2018} and might promote the fixation of traits that would get lost in uniform environments \citep{Krieger2020,Benzi2021,Plummer2019}. In wave-exposed coastal populations, such as mussel or seagrass beds, environmental variability due to cyclic tides and marine currents shapes pattern structure and makes these populations more resistant to long-term environmental shifts \citep{van2023optimal, Liu2014}. Environmental flows, therefore, play an important role both in determining population spatial structure and their long-term dynamics.

Most studies on the influence of flows on pattern formation in ecological systems have considered individual-based models in which the population is described by a collection of interacting particles \cite{Young2001} or continuous models based on partial differential equations for a population density field in which patterns form due to Turing or differential-flow instabilities \citep{Vasquez2008, Stucchi2013}. Nonlocal models, in which interactions between organisms are encoded in kernel functions that couple the dynamics of the population density across distant locations \cite{Pal2025}, have been barely analyzed in the presence of external flows. The nonlocal logistic equation (or nonlocal Fisher-Kolmogorov-Petrovsky-Piskunov, FKPP, equation), which assumes only diffusive dispersal and exponential population growth limited by long-range intraspecific competition, is one of the simplest models exhibiting pattern formation via nonlocal interactions \citep{Fuentes2003, Berestycki2009, Scheffer2006, Maruvka2006, Pigolotti2007}. With this minimum set of processes, patterns of population density can form in the low-diffusion regime if the intensity of the intraspecific competition between two organisms decays sharply enough with the distance between them (or, in mathematical terms, if the Fourier transform of the kernel function that defines this dependence of the competition strength with the distance between individuals is negative for some wavenumbers \citep{Hernandez-Garcia2004,LOPEZ2004223}). In these conditions, nonlocal interactions destabilize the uniform population distribution and lead to the self-organization of the system into a periodic pattern of spots of high population density, isolated by empty regions subject to very intense competition with the surrounding spots \citep{Hernandez-Garcia2004,LOPEZ2004223, Martinez-Garcia2013}. Because of the simple mechanism it proposes to explain pattern formation, the nonlocal FKPP equation has been the basis for developing more elaborated pattern-formation models testing the effect of additional processes on spatiotemporal population dynamics, such as demographic fluctuations or environmental heterogeneity \citep{Volpert2009, dornelas2019, dasilva2014, Jorge2024}.

Among these extensions, previous work coupled a chaotic flow to an individual-based formulation of the nonlocal logistic model \cite{Hernandez-Garcia2004}. Keeping all other demographic parameters constant, this study found that increasing mixing initially breaks the spotted pattern into non-stationary filaments, leading the population to a well-mixed configuration if mixing continues to increase. Across this gradient of mixing intensity, the stationary population size decreases monotonically. However, these results are obtained only through numerical simulations, because the complexity of the chaotic flow and the discreteness of the model make any analytical treatment impossible. Consequently, we still do not understand how environmental flows featuring different space-dependent velocity fields interact with spatial ecological feedbacks to determine pattern formation, population dynamics and stability. To fill this gap, we work with different extensions of a continuous nonlocal logistic model and focus on steady flows. These choices, which neglect demographic fluctuations and consider simpler shear and transport dynamics, allow us to go beyond numerical simulations and, in certain limits, develop analytical approximations to isolate the key effects of flows on the pattern formation instability.

We consider two flows. {First, a sine flow, whose simple velocity field serves as a toy model for shear flows in confined channels. This flow provides a simplified representation of, for example, the biophysical environment experienced by microbial populations in pipes where they are subject to anisotropic advective transport and shear forces \citep{Uppal2018}}. Second, we study a vortex flow with closed streamlines and shear forces that are not aligned with any of the primitive vectors of the population pattern \citep{Chandrasekhar2013,Zahnow2009,Abel2002}. For each of these velocity fields, we measure how the flow changes the conditions for pattern formation, the emergent spatiotemporal dynamics, and how these changes impact long-term population abundance. {One of our findings is that stationary velocity fields with a spatial structure that can not be aligned with the population density patterns that appear naturally in the absence of flow can induce a mixing effect and delay the pattern formation instability. Additionally, in the pattern formation regime, the pattern spatial structure and the flow velocity field jointly lead to a variety of population dynamics. These dynamics range from oscillations with a well-defined frequency for weak flow and population patterns aligned with the main direction of shear and transport, to seemingly chaotic dynamics when the pattern and the flow have very different spatial structures.}

\section{Materials and Methods}

\subsection{Flow models}\label{sec:flows}
We consider two stationary two-dimensional flows in a system of size $L \times L$ and 
periodic boundary conditions. This choice of environmental flows allows us to investigate how different spatial forms of fluid strain distort the emergent pattern of population density and how these changes in the pattern impact the long-term population size. The two flows 
are:

\noindent \textbf{Sine flow.} 
We first consider a sine flow in which the intensity of the $y$ velocity component is zero and the intensity of $x$ component changes along the $y$ axis. The velocity field is written as $v_0 {\boldsymbol{f}}({\bf x})$, with $v_0$ a typical velocity scale and ${\boldsymbol{f}}({\bf x})$ a dimensionless vectorial function with components
\begin{eqnarray}
f_x(x,y) &=& \sin \bigg(\frac{2\pi m}{L}y\bigg) \label{eq:vx_sin},\\
f_y(x,y) &=& 0 \label{eq:vy_sin},
\end{eqnarray}
where $m$ is a positive integer number that sets the number of flow periods in the system (see Panel A in Fig.\,\ref{fig:flowvector} for the vector field generated by this flow). We used $m=1$ in all the analyses. For this simple shear flow, the shear rate (proportional to the derivative of $f_x$ with respect to $y$) oscillates in the $y$ coordinate in anti-phase with the velocity.

\noindent\textbf{Rankine vortex flow.}
Second, we use a Rankine velocity field with radial symmetry \cite{Aref1983}, which we modify to make it suitable for the use of periodic boundary conditions \citep{Krieger2020,Miranda2023}. The Rankine vortex is a classic example of an irrotational vortex flow that corrects the singularity of the velocity at the vortex center by assuming that, in this region, the flow motion resembles the rotation of a solid body. In polar coordinates, the radial component of the velocity $v_r$ is zero, and the angular component depends on the distance to the center of the vortex as
\begin{equation} \label{eq:pointvortex-theta}
v_\theta(r,\theta) =\begin{dcases}
\frac{\Gamma r}{a^2} \quad \mbox{if} \quad  0\leq r < a,\\
\frac{\Gamma }{r}\quad \mbox{if} \quad  r\geq a
\end{dcases}
\end{equation}
where $r$ is the radial distance to the center of the vortex, which we placed at the center of the simulation domain; $a$ is the distance at which the intensity of flow reaches its maximum; and $\Gamma$ is the maximum vortex circulation. This parameter also defines the vortex spinning direction: counterclockwise for positive $\Gamma$ and clockwise for $\Gamma$ negative.

Because the vortex velocity field defined by Eq.\,\eqref{eq:pointvortex-theta} has an infinite range, considering periodic boundary conditions leads to a velocity field involving two infinite sums (see \citep{Miranda2023} for a detailed derivation using the point vortex model, which is also valid for the Rankine vortex if $a\ll L$). We can reduce the velocity field created by a periodic array of vortices to an expression in Cartesian coordinates that only uses one summation \citep{Weiss1991,Krieger2020,Miranda2023}
\begin{eqnarray}
v_x(\Delta x,\Delta y) &=& -\frac{\Gamma}{2L} \sum_{m=-\infty}^{\infty} \frac{\sin\left(\frac{2\pi \Delta y}{L}\right)}{\cosh\left[ 2\pi \left( \frac{\Delta x}{L} + m \right)\right]-\cos\left(\frac{2\pi \Delta y}{L}\right)}\label{eq:pointvortex-x}, \\
v_y(\Delta x,\Delta y) &=& \frac{\Gamma}{2L} \sum_{m=-\infty}^{\infty} \frac{\sin\left(\frac{2\pi \Delta x}{L}\right)}{\cosh\left[ 2\pi \left( \frac{\Delta y}{L} + m \right)\right]-\cos\left(\frac{2\pi \Delta x}{L}\right)},\label{eq:pointvortex-y}
\end{eqnarray}
where $\Delta x = x - L/2$ and $\Delta y = y - L/2$. {Eqs.\,\eqref{eq:pointvortex-x}-\eqref{eq:pointvortex-y} involve infinite sums over the index $m$ that can be treated with Ewald summation or similar techniques. However, the terms decay very fast with $m$, such that terms with $\rvert m \rvert = 2$ are $\mathcal{O}(10^{-20})$ \cite{Krieger2020}. We therefore truncate the series retaining only terms with $\rvert m \rvert \leq 1$}. Finally, we can define the characteristic velocity of this vortex flow $v_0=\Gamma/2L$, which will be used later to define a characteristic Péclet number. In all the simulations we performed using this periodic Rankine vortex flow, we used small values of $a$, such that the vortex radius is smaller than the range of the nonlocal ecological interactions.

\subsection{Nonlocal extensions of the logistic model}
{We study two modifications of the classical logistic model that are well established in the population dynamics literature. First, we consider a nonlocal extension of the FKPP in which spot patterns form due to long-range competition \cite{Fuentes2003}. Then, we focus on a more complex model in which organisms also interact via facilitative interactions, which leads to a variety of gap, labyrinth, and spot patterns \cite{Tlidi2008}. 
In both cases, we define a normalized population abundance,
\begin{equation}
    A(t) = \frac{1}{\rho^* L^2} \int_{S}  \rho({\bf x},t) \,\mathrm{d}{\bf x},
    \label{eq:At}
\end{equation}
where $\rho({\bf x},t)$ is the biomass density field, and $\rho^*$ is the density value of the stationary and homogeneous solution of the corresponding population-dynamics model. Therefore, $\rho^* L^2$ represents the steady-state total abundance in the absence of spatial patterns, and $A(t)$ quantifies the relative enhancement or depletion of the total abundance due to flow, compared to the no-flow case. When appropriate, we average this quantity over time once the system has reached a quasi-stationary regime to obtain a time-independent estimation of this population-abundance variation.} 

\subsubsection{The nonlocal Fisher-Kolmogorov-Petrovsky-Piskunov equation}\label{sec:model}

The nonlocal FKPP equation is an extension of the logistic model that describes the spatiotemporal dynamics of a single species diffusing in space and subjected to density-independent growth (asexual reproduction) limited by long-range intraspecific competition. Under these assumptions, the dynamics of the population density is given by
\begin{equation}
     \label{eq:Fkkp}
    \frac{\partial \rho({\bf x},t)}{\partial t} = \alpha \rho({\bf x},t) -\beta \rho({\bf x},t)\int_S G({\bf x-y})\rho({\bf y},t) \,\mathrm{d}{\bf y} + D\boldsymbol{\nabla}^2_x  \rho({\bf x},t),
\end{equation}
where $D$ and $\alpha$ are the diffusion constant and the net linear growth rate, respectively. $\beta$ modulates the intensity of intraspecific competition, $S$ indicates that the integral is performed over the entire simulation domain. In all our analyses, we use a square domain of lateral length $L$  with periodic boundary conditions, which is the standard choice in spatial models of population dynamics. This choice of boundary conditions turns the integration domain into a $2D$ torus. 
$G({\bf x}-{\bf y})$ is the competition kernel that defines how strongly the death rate of organisms at a focal location ${\bf x}$ is influenced by the presence of conspecifics at neighboring locations ${\bf y}$. We consider that the intensity of nonlocal competition scales with population density (rather than population abundance), which implies that the competition kernel must be normalized to one. This condition guarantees that Eq.\,\eqref{eq:Fkkp} becomes the logistic model 
if $\rho({\bf x},t)$ is uniform in space.

Previous work has shown that Eq.\,\eqref{eq:Fkkp} can exhibit spatially periodic patterns of population density when the Fourier transform of $G({\bf x-y})$ takes negative values for some wavenumbers \citep{Hernandez-Garcia2004,LOPEZ2004223}. As an example of this situation, we assume that the process underlying long-range competition is isotropic and that two organisms at a distance $\rvert {\bf x}-{\bf y}\rvert$ will compete with each other with a distance-independent strength provided they are closer than a threshold distance $R$. Mathematically, these assumptions imply choosing an isotropic top-hat kernel of the form
\begin{equation}\label{eq:kernel}
    G( {\bf x} ) =
    \begin{dcases}
        \frac{1}{\pi R^2} & \mbox{if} \ \rvert{\bf x}\rvert\leq R, \\
        0 & \mbox{otherwise}. \\
    \end{dcases}
\end{equation}

We extend the nonlocal FKPP equation in \eqref{eq:Fkkp} to introduce the effect of transport by a stationary flow. For these flows, the velocity field does not change in time and therefore it can be written as ${\bm v}({\bf x},\,t)=v_0{\bm f}({\bf x})$, where $v_0$ is a typical velocity and ${\bm f}({\bf x})$ is a vector dimensionless function giving the spatial functional dependence of the velocity field.
Under the additional assumption that the velocity field is incompressible, this additional transport process leads to a model equation
\begin{equation}
    \label{eq:Fkkp_advection}
    \frac{\partial \rho({\bf x},t)}{\partial t} = \alpha \rho({\bf x},t) -\beta \rho({\bf x},t)\int_S G({\bf x-y})\rho({\bf y},t) \,\mathrm{d}{\bf y} + D\boldsymbol{\nabla}^2_{\bf x}  \rho({\bf x},t) - v_0{\bm f}({\bf x})\cdot \boldsymbol{\nabla}_{\bf x}\rho ({\bf x},t),
\end{equation}
where the kernel is still given by Eq.\,\eqref{eq:kernel}, assuming that the spatial dependence of the long-range interaction does not change because of the flow. 

To perform analytical calculations in Eq.\,\eqref{eq:Fkkp_advection}, we obtain its dimensionless version using the following scaled quantities,
\begin{equation}
    \label{eq:units_rescale}
    {\bf x} \rightarrow R{\bf u}; \quad t \rightarrow \frac{\tau R^2}{D}; \quad \rho({\bf x},t) \rightarrow \frac{D}{R^{2} \beta} {\rho}({\bf u},\tau); \quad
    G({\bf x}) \rightarrow \frac{H({\bf u})}{R^2}\, ,
\end{equation}
where the scaled kernel remains normalized, $\int_E H({\bf u}) d{\bf u} =1$ and we will denote the Cartesian components of $\bf u$ as $(u,v)$. Using these scaled variables, we get
\begin{equation}\label{eq:FKKPscaled}
    \frac{\partial \rho({\bf u},\tau)}{\partial \tau} = \mathrm{Da} \,\rho({\bf u},\tau) + \boldsymbol{\nabla}^2_u  \rho({\bf u},\tau) - \rho({\bf u},\tau)\int_E H({\bf u-v})\rho({\bf v},t) \,\mathrm{d}{\bf v} - \mathrm{Pe}\,\boldsymbol{f}({\bf u})\cdot \boldsymbol{\nabla}_u\rho ({\bf u},\tau)\,,
\end{equation}
where $E$ is the scaled domain, with lateral length $L/R$ and still with periodic boundary conditions. We have, moreover, defined two dimensionless control parameters
\begin{equation}\label{eq:Pec-def}
 \mathrm{Da} = \frac{\alpha R^2}{D} \;\;\;\; \text{and} \;\;\;\; \mathrm{Pe} = \frac{v_0R}{D}\,,
\end{equation}
that are, respectively, the diffusive Damköhler number and a characteristic Péclet number. Finally, in the spatial structure of the velocity field, we have simply replaced ${\bm f}({\bf x}) \rightarrow {\bm f}({\bf u})$ because it is already dimensionless.

\subsubsection{The logistic model with nonlocal competition and facilitation}\label{sec:vegetation}

We analyze a second extension of the logistic equation that also accounts for nonlocal positive intraspecific interactions.
In this model, the dynamics of the biomass density is given by
\begin{align}\label{eq:sdf}
    \frac{\partial \rho({\bf x},t)}{\partial t} = b\,\mathcal{M}_\mathrm{f}\lbrack{\rho}\rbrack \rho({\bf x},t) \left( 1 - \frac{\rho({\bf x},t)}{K}\right) -\mu \mathcal{M}_\mathrm{c} \lbrack{\rho}\rbrack \rho({\bf x},t) + D\boldsymbol{\nabla}^2_{\bf x} \rho({\bf x},t),
\end{align}
where $K$, $b$, and $\mu$ are non-negative constants representing the carrying capacity, baseline birth and death rates in the absence of nonlocal interactions. $\mathcal{M}_{\mathrm{f, c}} \lbrack{\rho}\rbrack $ are birth and death rate enhancement factors due to nonlocal interactions. Following the original model formulation\cite{Tlidi2008}, these enhancement factors are defined as
	\begin{align}\label{eq:defcoeff}
		\mathcal{M}_{\mathrm{f, c}} \lbrack{\rho}\rbrack  = \exp \bigg( \chi_{\mathrm{f, c}} \int G_{\mathrm{f, c}}({\bf x}-{\bf y}) \rho({\bf y},t) \,\mathrm{d}{\bf y}\bigg).
	\end{align}
where $\chi_{\mathrm{f, c}}$ are non-negative constants modulating how strongly long-range interactions impact the population dynamics and $G_{\mathrm{f, c}}$ are the interaction kernels. Also following the original model formulation in \cite{Tlidi2008}, we use isotropic Gaussian kernels given by
	\begin{align}
		G_{\mathrm{f, c}}({\bf x}-{\bf y}) =\mathcal{N} e^{-({\bf x}-{\bf y})^2 /R^2_{\mathrm{f, c}} },
	\end{align}
where $\mathcal{N} = 1/\pi R^2_{\mathrm{f, c}}$ is a normalization factor and $R^2_{\mathrm{f, c}}$ sets the characteristic spatial scale of positive and negative feedbacks. Note that Eq.\,\eqref{eq:defcoeff} imposes $\mathcal{M}_{\mathrm{f, c}}\geq 1$, and that the enhancement factors are equal to one when the modulating coefficients $\chi_{\mathrm{f, c}}$ are equal to zero. 

We can also extend Eq.\,\eqref{eq:sdf} to introduce the effect of transport by a stationary flow,
\begin{align}\label{eq:veg_model}
    \frac{\partial \rho({\bf x},t)}{\partial t} = b\,\mathcal{M}_f \lbrack{\rho}\rbrack  \rho({\bf x},t)\left( 1 - \frac{\rho({\bf x},t)}{K}\right) -\mu \mathcal{M}_c \lbrack{\rho}\rbrack \rho({\bf x},t) + D\boldsymbol{\nabla}^2_{\bf x}\rho({\bf x},t )- v_0\,{\bm f}({\bf x})\cdot \boldsymbol{\nabla}_{\bf x}\rho ({\bf x},t),
\end{align}
where we can scale space, time, and density as
\begin{equation}
    \label{eq:units_rescale2}
    {\bf x} \rightarrow R_\mathrm{c} {\bf u}; \quad t \rightarrow \frac{R_\mathrm{c}^2}{D}\tau; \quad \rho({\bf x},t) \rightarrow \frac{{\rho}({\bf u},\tau)}{K}; \quad G_{\mathrm{f,c}}({\bf x})\rightarrow \frac{H_{\mathrm{f,c}}({\bf u})}{R_\mathrm{c}^2}
\end{equation}
to obtain a dimensionless equation
\begin{equation}
     \frac{\partial \rho({\bf u},\tau)}{\partial \tau} = 
 \mathrm{Da}^+\mathcal{M}_f\lbrack \rho \rbrack \rho({\bf u},\tau)\big[1-\rho({\bf u},\tau)\big] - \mathrm{Da}^-\mathcal{M}_c\lbrack \rho \rbrack \rho({\bf u},\tau) + \boldsymbol{\nabla}^2_{\bf u}\rho({\bf u},\tau)- \mathrm{Pe} \,{\bm f}({\bf u})\cdot \boldsymbol{\nabla}_{\bf u}\rho ({\bf u},\tau), \label{eq:vegetation_escaled}
\end{equation}
with 
\begin{equation} \label{eq:Pec-def2}
 \mathrm{Da}^+ = \frac{b R_c^2}{D}; \quad \mathrm{Da}^- = \frac{\mu R_c^2}{D}; \quad \mathrm{Pe} = \frac{v_0 R_c}{D}; \quad \overline{\chi}_{\mathrm{f, c}}= \frac{\chi_{\mathrm{f, c}}}{K} \,, 
\end{equation}
and the factors $\mathcal{M}_{\mathrm{f,c}}$ are already computed using scaled variables and kernels.
\color{black}
    
\section{Results}\label{sec:results}

For each of the models we considered, we first study the no-flow limit $\mathrm{Pe}=0$ to use it as a baseline for quantifying the effect of advection on both the spatial patterns of population density and the long-term population size. We study this no-flow limit analytically via a linear stability analysis and perform numerical simulations of the fully nonlinear equations (see App.\,\ref{app:numerical} for details of the pseudospectral methods we used and \href{https://github.com/Jaegg3rNat/NonLocalShear.git}{GitHub} for the code repository). The linear stability analysis is a standard technique in pattern-formation theory that consists on adding small spatial perturbations to one of the uniform solutions of the fully nonlinear model and obtaining the short-term linearized perturbation growth rate \citep{Cross1993}. If this perturbation growth rate is positive for a certain wavenumber, then perturbations with that periodicity will grow over time, and the system may exhibit spatial patterns. Otherwise, the perturbation decays, meaning that the uniform solution is stable against small spatial perturbations.

\subsection{The nonlocal FKPP} \label{sec:noflow}

The nonlocal FKPP equation \eqref{eq:Fkkp} has a trivial solution $\rho=0$, which is the only non-negative homogeneous steady solution when $\mathrm{Da}<0$. For $\mathrm{Da}>0$, a new homogeneous steady solution appears, $\rho^*=\mathrm{Da}$. To perform a linear stability analysis on this solution, we consider a perturbed solution of the form $\rho({\bf u},\tau) = \rho^* + \epsilon\psi({\bf u},\tau)$ where $ \rho^* = \mathrm{Da}$ is the equilibrium uniform solution, and $\epsilon\ll 1$. The linearized dynamics of the perturbation is given by
\begin{equation}\label{eq:fluc_dynamics}
      \frac{\partial \psi({\bf u},\tau)}{\partial \tau} =  \boldsymbol{\nabla}^2_u  \psi({\bf u},\tau) -\,\mathrm{Da} \int_E H({\bf u-v})\psi({\bf v},\tau) \,\mathrm{d}\bf{v},
\end{equation}
which is a linear equation that we can solve using the Fourier transform. Fourier transforming Eq.\,\eqref{eq:fluc_dynamics}, and considering that the transformed equation is linear, we can assume for the Fourier transform of the perturbation that $\hat{\psi}(\bm{k},\tau)\propto \exp\left(-\lambda(\bm{k})\tau\right)$. Note that, because we are using scaled variables, the wavevectors ${\bm{k}}$ are measured in units of $R^{-1}$.

\begin{figure}[H]
    \centering
    \includegraphics[width=0.6\textwidth]{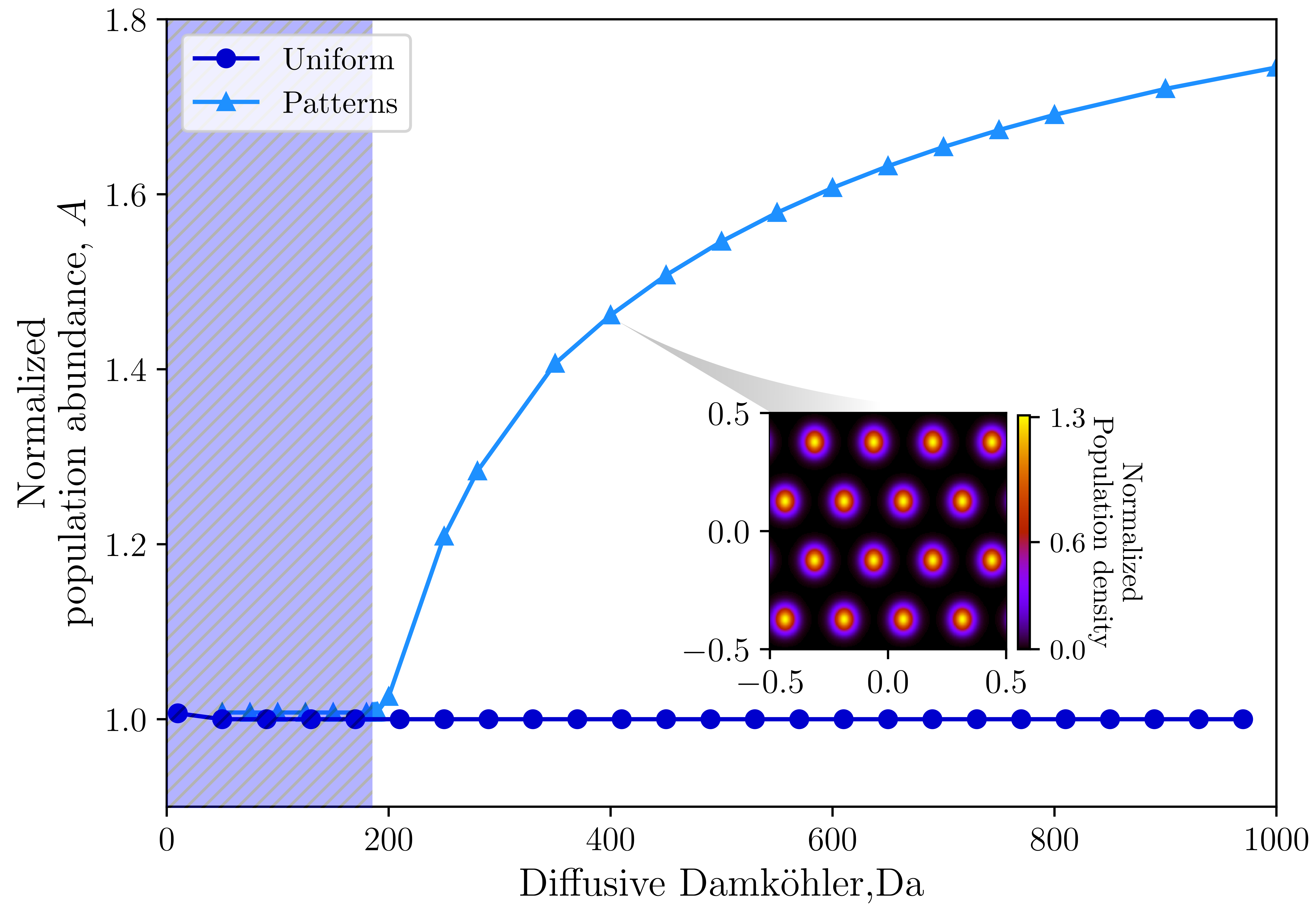}
    \caption{Normalized population abundance ($A(t)$, from Eq.\eqref{eq:At}), time-averaged at long times, as a function of $\mathrm{Da}$ for the patterned and uniform distributions of population density for the nonlocal FKPP equation in the absence of flow. The blue-shaded region highlights the non-pattern parameter regime where $\mathrm{Da}<\mathrm{Da}_\mathrm{c}\approx 185.192$. The inset shows a typical pattern of the normalized population density as a function of the unscaled spatial variables ${\bf x}=(x,y)$ for
 $\mathrm{Da}>\mathrm{Da}_\mathrm{c}$.}
    \label{fig:1}
\end{figure}

We obtain an equation for the perturbation growth rate
\begin{equation}
    \lambda(k) = -k^2- \mathrm{Da}\,\hat{H}(k)
    \label{eq:dispersion1}
\end{equation}
where $\hat{H}(k)$ is the Fourier transform of the scaled competition kernel, and $k=\,\rvert {\bm k} \rvert$ is the modulus of the wavevector ${\bm k}$. Because the first term on the right side of Eq.\,\eqref{eq:dispersion1} is always negative and $\mathrm{Da}$ is a positive constant, pattern formation requires that the Fourier transform of the competition kernel must be negative for some wavenumber $k$ \citep{Hernandez-Garcia2004,LOPEZ2004223,Pigolotti2007}. {In general, if we consider a family of kernels defined by generalized Gaussian functions, $G(x)\propto \exp\left[-{(\rvert x \rvert / R)}^p\right]$ with $p>0$, pattern formation will be possible for $p>2$. The scaled top-hat kernel, $H({\bf u})$ we introduced in Eqs.\,\eqref{eq:kernel} and \eqref{eq:units_rescale} corresponds to the $p\rightarrow \infty$ limit within this family of kernels. Its Fourier transform is}
\begin{equation}
    \hat{H}(k) =  2 \frac{J_1( k )}{k },
\end{equation}
where $J_1$ is the order-1 Bessel function of the first kind and thus an oscillatory function taking negative values for some wavevector modulus $k$. Therefore, when $\mathrm{Da}$ is large enough, $\mathrm{Da}>\mathrm{Da}_{\mathrm{c}}\approx 185.192$, the uniform distribution of population density becomes unstable and patterns form with critical wavenumber $k_c \approx 4.79$. In the numerical simulations of the fully nonlinear equation, we observe that these patterns have a hexagonal symmetry. This spatial configuration minimizes competition throughout the entire system and therefore leads to population abundances higher than in the uniform case (Fig.\,\ref{fig:1}) \cite{Hernandez-Garcia2004}. Consequently, in the nonlocal FKPP, spatial patterns are a signature of high productivity, emerging only at high reproduction rates (\textit{i.e.} high $\mathrm{Da}$).

Next, we investigate how advection by the flows introduced in Section \ref{sec:flows} alters the patterns formed due to long-range interactions and how these new spatial configurations of population density affect population abundance. We perform these analyses by combining numerical simulations and, when possible, analytical approximations. For the numerical simulations, we used the dimensional equation \eqref{eq:Fkkp_advection} with $D =10^{-4}, \beta =1, R=0.2L$ and $L=1$ fixed throughout all the analyses and varying other parameters.

\subsubsection{Sine flow}

To get a first understanding of how the sine flow might impact pattern formation, we extend the linear stability analysis of Section \ref{sec:noflow} to account for an advection term with a velocity field given by Eqs.\,\eqref{eq:vx_sin}-\eqref{eq:vy_sin} with $m=1$ (see App.\,\ref{app:LSA-flow} for the details of the calculation). We performed a linear stability analysis on the nonlocal FKPP to simplify our calculations but we will show numerically that the results also hold for the logistic model with nonlocal competition and facilitation (Section \ref{sec:SDFresults}). The linearized dynamics of a perturbation to the uniform solution is now given by
\begin{equation}\label{eq:linear-flow}
      \frac{\partial \psi({\bf u},\tau)}{\partial \tau} =  \boldsymbol{\nabla}^2_u  \psi({\bf u},\tau) - \,\mathrm{Da} \int_E H({\bf u-v})\psi({\bf v},\tau) \,\mathrm{d}{\bf v}- \, \mathrm{Pe}\, \sin\left(\tilde{q}\, v\right)\frac{\partial \psi({\bf u},\tau)}{\partial u}.
\end{equation}
where $\tilde q=2 \pi/L$ and $L$ is measured in units of $R$. Because the system is periodic, we can use a Floquet decomposition of the perturbation that gives a general expression for the perturbation growth rate $\lambda({\bf k})$
\begin{equation}\label{eq:flowinfinity}
    \left[\lambda({\bf k}) + k_n^2 + \mathrm{Da} \hat{H}({\bf k}_n)\right] C_n({\bf k}) = k_u \frac{\mathrm{Pe}}{2} \big[C_{n+1}({\bf k})-C_{n-1}({\bf k})\big] \quad n = 0,1,2...
\end{equation}
where $C_n({\bf k})$ are Fourier-Floquet coefficients of the perturbation and we have defined ${\bf k}_n=(k_u,k_v+n\tilde{q})$ for $n = 0,1,2...$ (see App.\,\ref{app:LSA-flow}). Note also that, for $\mathrm{Pe}=0$ or $k_u=0$, Eq.\,\eqref{eq:flowinfinity} for $n=0$ reduces to the expression for the perturbation growth rate we derived without flow, Eq.\,\eqref{eq:dispersion1}. {Thus, the value of $\lambda(k_u=0,k_v)$ is not affected by the flow and it will become positive at the same value of $\mathrm{Da} = \mathrm{Da}_c$ as in the absence of flow. Consequently, when increasing the value of $\mathrm{Da}$ in the presence of the flow, the homogeneous population will become unstable no later than at $\mathrm{Da} =\mathrm{Da}_c$, giving rise to a pattern of wavenumber ${\bf k} = (k_u = 0,k_v)$, which corresponds to horizontal stripes (\textit{i.e.}, aligned along the $u$ direction).} We have to investigate if other types
of instabilities appear at lower values of $\mathrm{Da}$ or if this is the first instability of the homogeneous solution encountered when increasing $\mathrm{Da}$. To this end, in the $\mathrm{Pe}\ll 1$ limit, we can truncate the system of equations in Eq.\,\eqref{eq:flowinfinity} to only consider $n=0,\,\pm 1$ (see Appendix \ref{app:LSA-flow}). This type of system has a non-trivial solution only if the determinant of the coefficient matrix is zero, which gives us three eigenvalues $\lambda_0({\bm k})$, and $\lambda_{\pm}({\bm k})$. By numerically analyzing the behavior of these eigenvalues, we find that the real part of the largest eigenvalue becomes positive at the same value $\mathrm{Da}=\mathrm{Da}_\mathbf{c}$ as in the absence of flow, indicating that the sine flow does not play any role in setting the onset of pattern formation (Fig.\,\ref{fig:eigen}). However, the characteristics of the spatial patterns are strongly affected by the flow, as described in the following.

\begin{figure}
    \centering
    \includegraphics[width=0.8\textwidth]{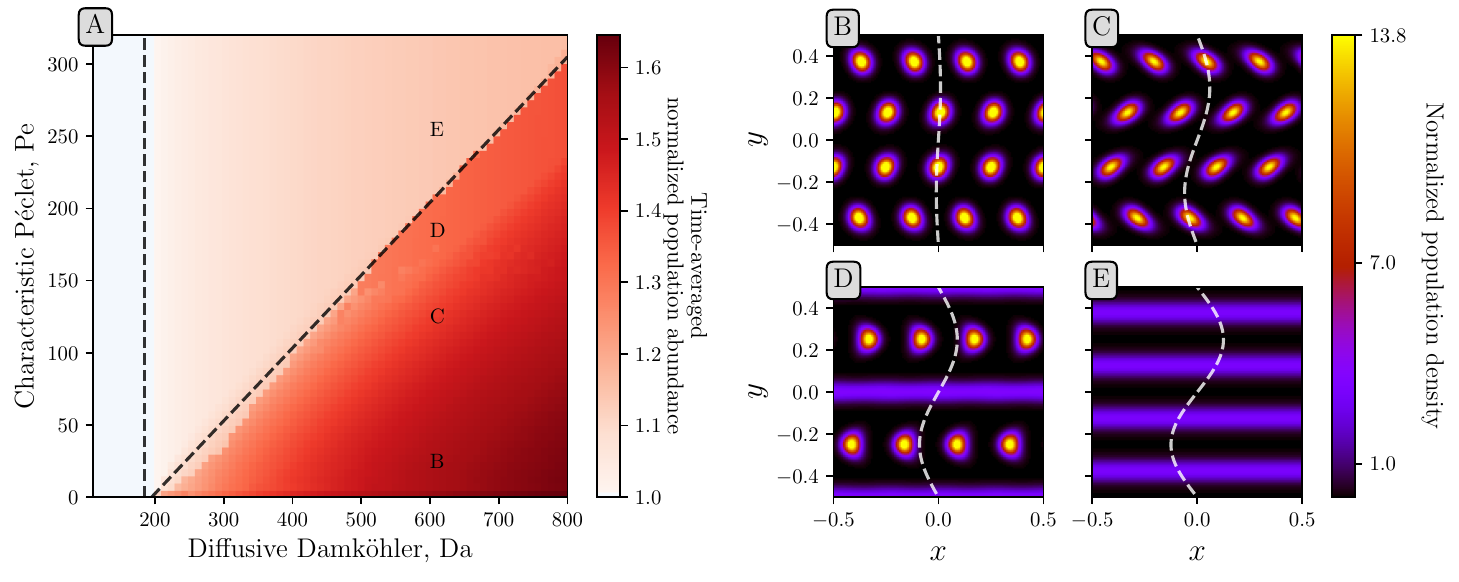}
    \caption{A) Time-averaged normalized population abundance as a function of $\mathrm{Da}$ and $\mathrm{Pe}$ for the nonlocal FKPP equation in the presence of the sine flow. We performed the time average for $t>2000$ and sampled the spatial configuration until the relative standard error of this temporal average was below $1\%$. The black dashed lines limit the regions with the different spatial patterns shown in (B-E), and they are computed from the maxima in the modulus of the gradient in this panel (see Fig.\,\ref{fig:SM1}). Darker red represents higher abundance relative to the uniform case (white). B-E) Snapshots of the normalized density spatial pattern once the system reaches an asymptotic state. We find four possible spatiotemporal patterns with increasing flow intensity: (B) spots moving horizontally ($\mathrm{Da} = 600, \mathrm{Pe} = 20$), (C) stretched moving spots ($\mathrm{Da} = 600,  \mathrm{Pe} = 120$), (D) coexisting spots and stripes ($\mathrm{Da} = 600, \mathrm{Pe} = 180$), and (E) stripes ($\mathrm{Da} = 600, \mathrm{Pe} = 250$) (see \href{https://www.dropbox.com/scl/fo/ffev9lhkfshavi98jo7ck/AGZlO-xaIipWSAQUzjIe4UA?rlkey=hskk32s6ehj9gdvfwdfrdhafa&dl=0}{SM Videos}).}
    \label{fig:heat_map}
\end{figure}

Numerical simulations of the full, nonlinear equation confirm the theoretical prediction that the sine flow does not change the value of the Damköhler number at which the uniform solution loses stability and patterns form (Fig.\,\ref{fig:heat_map}A, Fig.\,\ref{fig:SM1}). Increasing $\mathrm{Da}$ at any fixed value of ${\mathrm{Pe}}$, the first instability appearing on the homogeneous state leads to horizontal stripes (Fig.\,\ref{fig:heat_map}E), in agreement with the linear stability analysis sketched above and in Appendix \ref{app:LSA-flow}. Moreover, the spot-stripe pattern transition only depends on the ratio ${\mathrm{Pe}}/\mathrm{Da}$ and is thus independent of the specific value of the diffusion constant provided that $D$ is low enough to allow for pattern formation in the absence of advection. When we explore the full $(\mathrm{Da}, {\mathrm{Pe}})$ parameter space, however, we find that the environmental flow leads to a diversity of outcomes, both in terms of the spatial patterns and the total population size. As expected from the analysis without flow, the population abundance increases with the Damköhler number $\mathrm{Da}$. Increasing the maximum flow intensity ${\mathrm{Pe}}$, however, reduces the total population abundance because the flow deforms the hexagonal pattern of circular spots (inset of Fig.\,\ref{fig:1} and Figs.\,\ref{fig:heat_map}B-E) and makes neighbor spots compete with each other.

The changes in total population size are not smooth over the entire parameter space. We can quantify these irregularities by computing the modulus of the gradient of the time-averaged normalized population abundance in the $\mathrm{Pe}$-$\mathrm{Da}$ plane (Fig.\,\ref{fig:SM1}). Non-smooth changes in the total population size as the Damköhler and Péclet numbers vary allow us to define at least three different regions within the parameter space (assuming $\mathrm{Da}>\mathrm{Da}_\mathrm{c}$; see Fig.\,\ref{fig:SM1}). In the first region, found when increasing $\mathrm{Pe} $ at fixed $\mathrm{Da}$, at the bottom right corner of Fig.\,\ref{fig:heat_map}A, the flow stretches and transports the population spots horizontally, making the pattern non-stationary and formed by elliptic spots (Figs.\,\ref{fig:heat_map}B,\,C and \href{https://www.dropbox.com/scl/fi/rygyr6awo0hfq2czkygy4/sine_mu600_pe20.mov?rlkey=p8187hvy20bd45d1s3b1c8fht&dl=0}{SM Video 1}, \href{https://www.dropbox.com/scl/fi/pz4yphqgutsj5mp8vuxqh/sine_mu600_pe120.mov?rlkey=y9at07opr3v2c8dwqknxx4s55&dl=0}{SM Video 2}). 
Within this region, higher values of ${\mathrm{Pe}}$ increase the stretch of the spots, which makes spots in the same pattern row interact more strongly with their neighbors.
As ${\mathrm{Pe}}$, and consequently shear, continues to increase, spots placed in the high-shear regions of the environment are fully stretched and become stripes (Fig.\,\ref{fig:heat_map}D; \href{https://www.dropbox.com/scl/fi/u9hgzjyr82y9h9mg9chdi/sine_mu600_pe180.mov?rlkey=oon8addkrnz29wf0rgrmwr6od&dl=0}{SM Video 3}). Finally, when the ratio between ${\mathrm{Pe}}$ and $\mathrm{Da}$ is high, the shear of the flow is strong enough to break the spots everywhere in the environment, and stripes aligned with the flow velocity form everywhere in the system (Fig.\,\ref{fig:heat_map}E; \href{https://www.dropbox.com/scl/fi/8i5zxvcaj4me9dct461xp/sine_mu600_pe250.mov?rlkey=mn0b9xst1ytk8ayiq0uznzux8&dl=0}{SM Video 4}). 

\begin{figure}
    \centering    \includegraphics[width=0.75\textwidth]{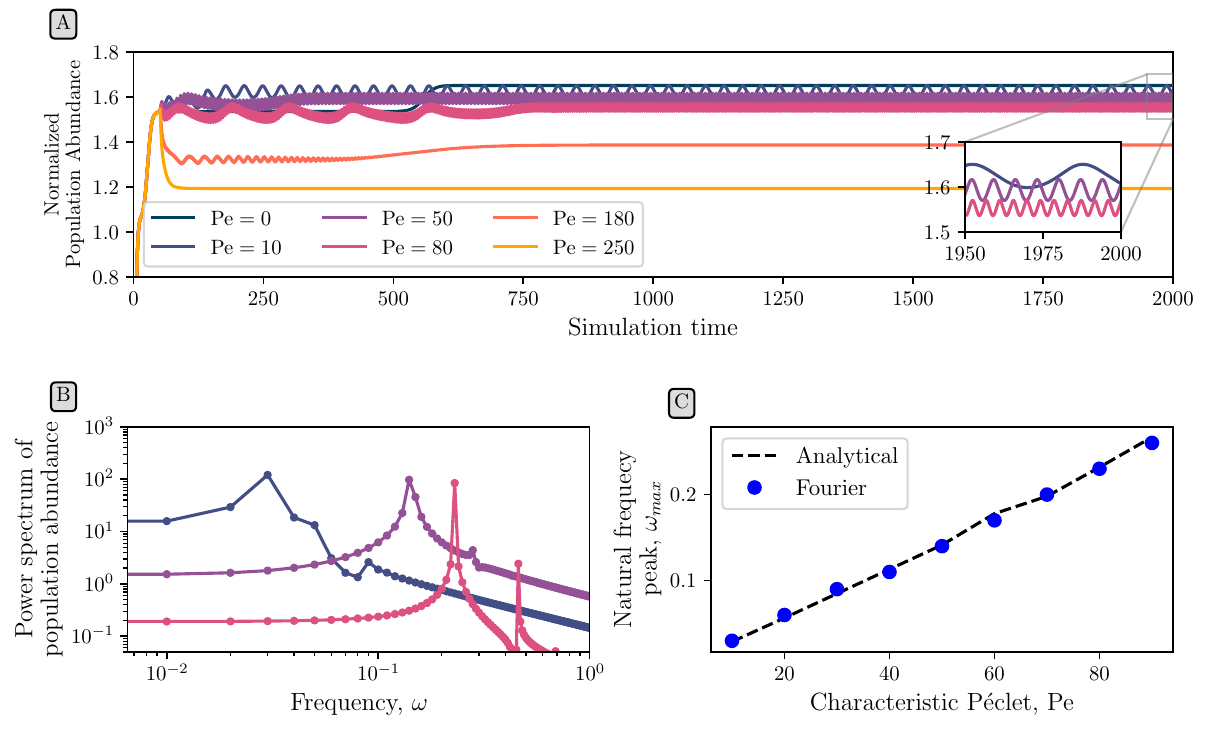}
    \caption{A) Time series of the normalized population abundance for the nonlocal FKPP equation with the sine flow. $\mathrm{Da} = 600$, and different values of $\mathrm{Pe}$. B) Log-log power spectrum of the time series in panel A at long times (between $t\in [800,2000]$. C) Natural frequency of the time series, defined as the frequency at which the power spectrum reaches its maximum value, as a function of $\mathrm{Pe}$. Blue dots correspond to the values extracted from the power spectra obtained from numerical simulations and the black dashed line shows the values obtained by approximating population-density spots by point-like particles centered at the spot center.}
    \label{fig:fourier_sine}
\end{figure}

Next, we looked at how the changes in the spatial pattern induced by the shear (transport and stretch of the population density spots) impact the time series of population abundance (Fig.\,\ref{fig:fourier_sine}). For low ${\mathrm{Pe}}$ (B and C regions), the total population abundance oscillates, and the frequency of these oscillations is determined by the wavelength of the pattern and the speed at which the environmental flow changes the symmetry of the pattern as it transports its spots. Without the flow, the spots have a hexagonal symmetry typical of competition-induced patterns. This geometry maximizes population size because it minimizes competition \citep{Hernandez-Garcia2004, Jorge2024}.

A weak sine flow, however, transports rows of spots in the $x$ direction, making them periodically interact with the spots in the upper and lower rows and inducing oscillations in the total population size. To characterize these oscillations quantitatively, we computed the power spectrum of the time series of total population abundance, which we obtained by integrating the spatial pattern over the system size. These power spectra show a peaked maximum at a non-zero frequency that corresponds with the natural frequency at which population size oscillates (Fig.\,\ref{fig:fourier_sine}B). To test whether the total population size oscillates due to the transport of the spots by the environmental flow as described previously, we calculate the frequency at which the pattern alternates its shape between a hexagonal and a square structure. This transition occurs while the flow transports rows of spots at a $y$-dependent velocity, and we can calculate its frequency by approximating spots by a point particle placed at the spot center and calculating the time this point particle needs to travel a distance equal to one-half of the pattern wavelength. When the flow is weak so that spots are not stretched, the results obtained from this mapping return an excellent agreement with the frequency extracted from the power spectra of the simulations, $\omega_\mathrm{max}$ (Fig.\,\ref{fig:fourier_sine}C). At higher ${\mathrm{Pe}}$, however, the flow stretches the spots, the pattern loses its hexagonal structure, and this approximation fails.

\subsubsection{Rankine vortex flow}

The results obtained with a sine flow suggest that the interplay between the spatial structure of the flow velocity field and the spatial pattern is key to determining population dynamics. To explore this connection more deeply, we next investigate a scenario with a Rankine vortex flow. Because this flow has circular symmetry, it can not be aligned with the hexagonal pattern that forms in the absence of advection, nor with any of its primitive vectors. For this flow, the linear stability analysis becomes too difficult and, thus, we describe in the following the results of numerical simulations.

\begin{figure}
    \centering
    \includegraphics[width=0.8\textwidth]{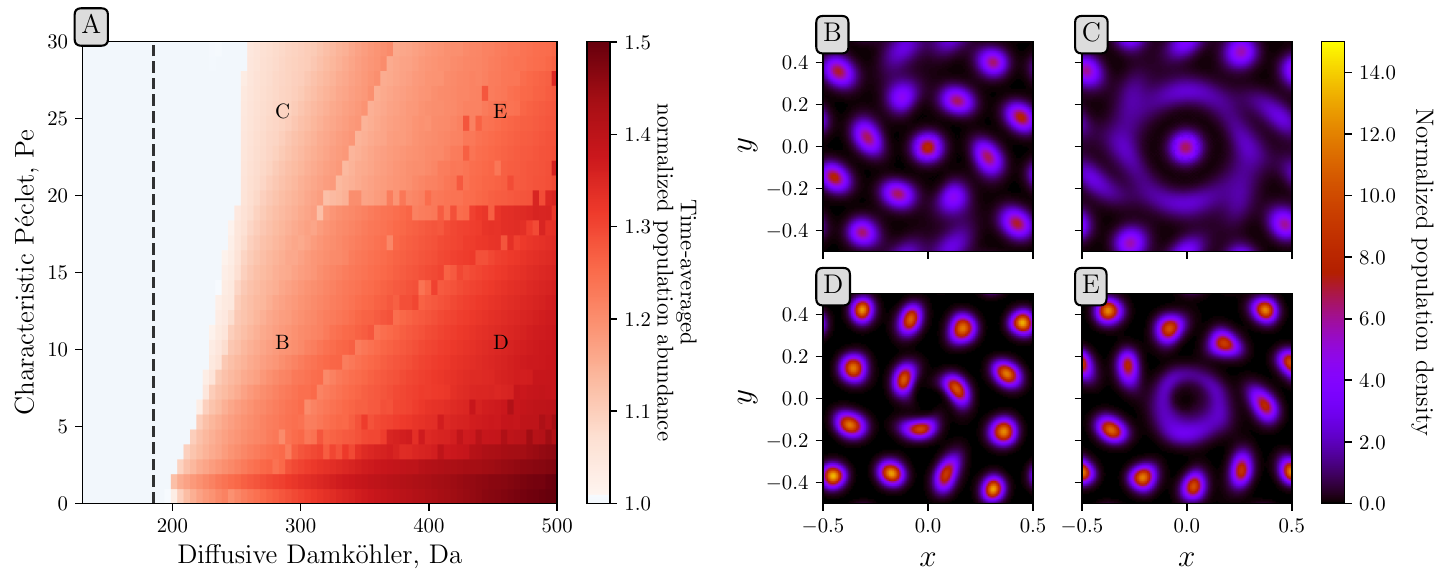}
    \caption{A) Time-averaged normalized population abundance as a function of $\mathrm{Da}$ and $\mathrm{Pe}$ for the nonlocal FKPP equation in the presence of the Rankine vortex flow. The black dashed line indicates the Damköhler number, $\mathrm{Da}_\mathrm{c}$, for the non-flow pattern instability. Darker red represents higher abundance relative to the uniform case (white). B-E) Snapshots of the long-term normalized density spatial patterns: (B) rotating spots ($\mathrm{Da} = 280,\, \mathrm{Pe} = 10$); (C) central spot surrounded by a ring of creation-annihilation spots ($\mathrm{Da} = 280, \,\mathrm{Pe} = 25$); (D) stretched rotating spots ($\mathrm{Da} = 450,\, \mathrm{Pe} = 10$); and (E) central ring surrounded by rotating spots ($\mathrm{Da} = 400,\,\mathrm{Pe} = 25$) (see SM Videos 5-8 for the pattern formation dynamics). Rankine vortex radius $a=0.05L$}
    \label{fig:heat_map_vortex}
\end{figure}

For this Rankine flow, the time-averaged normalized population abundance also decreases with increasing $\mathrm{Pe}$ and, as expected, increases with the Damköhler number $\mathrm{Da}$ (Fig.\,\ref{fig:heat_map_vortex}). The emergent spatiotemporal patterns of population density are, however, much more complex than those produced by the sine flow. First, and most important, we observe that the rotational velocity shifts the transition to patterns to higher values of the Damköhler number (Fig.\,\ref{fig:heat_map_vortex}). Moreover, the shift in $\mathrm{Da}_\mathrm{c}$ increases with $\mathrm{Pe}$ until, it appears, it plateaus for sufficiently large Péclet. Regarding pattern dynamics, the circular movement of spots caused by the radial symmetry of the velocity field makes them interact more with each other, including the possibility of spot-splitting annihilation dynamics that we did not observe for the sine flow.

Because the spots of population density undergo this more complex spatiotemporal dynamics, we cannot divide the parameter space into well-defined regions using a quantitative criterion as we did for the sine flow (see Fig.\,\ref{fig:SM2}). Instead, we systematically explore the parameter space and describe the different types of spatiotemporal patterns we observe. For low $\mathrm{Da}$ and low $\mathrm{Pe}$, the pattern rotates around the central spot that resides inside the region where the angular velocity increases linearly with $r$ (Fig.\,\ref{fig:heat_map_vortex}B and \href{https://www.dropbox.com/scl/fi/oxwlz9felbvasfv5oxorl/Rankine_video280_10.mov?rlkey=2hqlexwm7x9qt4y452pddnczg&dl=0}{SM Video 5}). When $\mathrm{Da}$ is still low, but $\mathrm{Pe}$ increases, the vortex origin is populated and surrounded by a ring of population density where spots are constantly absorbed and ejected (Fig.\,\ref{fig:heat_map_vortex}C; \href{https://www.dropbox.com/scl/fi/tfew04gchyif914qxugpn/Rankine_video280_25.mov?rlkey=dm0f6irxcdjs8i0dp50g9wa19&dl=0}{SM Video 6}). At larger values of $\mathrm{Da}$, the central region of the environment is occupied by three stretched spots that rotate in a direction given by the vortex circulation (Fig.\,\ref{fig:heat_map_vortex}D; \href{https://www.dropbox.com/scl/fi/m7mbs87evv5yqzyqw3aip/Rankine_video450_10.mov?rlkey=7lbk47z4e2yevtskxcokb3s2p&dl=0}{SM Video 7}). Within this high-$\mathrm{Da}$ regime, the pattern keeps the same qualitative organization we found in Fig.\,\ref{fig:heat_map_vortex}D when $\mathrm{Pe}$ increases. However, due to higher stretching in the spots, the gap between the stretched spots closes and these form a stable ring (Fig.\,\ref{fig:heat_map_vortex}E; \href{https://www.dropbox.com/scl/fi/bvysd3x0cjntnb6v8vtw0/Rankine_video450_20.mov?rlkey=4kwr3dvrz36sep4zh9devw3lc&dl=0}{SM Video 8})

\begin{figure}
    \centering
    \includegraphics[width=0.8\textwidth]{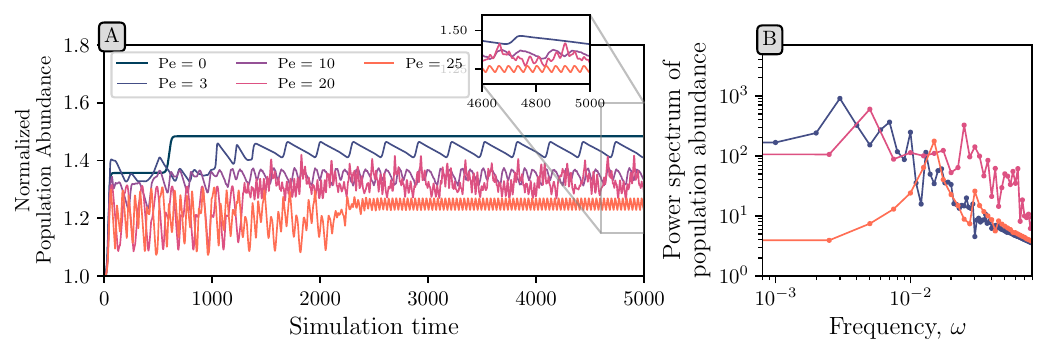}
    \caption{A) Time series of the normalized population abundance for the nonlocal FKPP eauation for the Rankine vortex flow. B) Log-log power spectrum of the time series in panel A in the long-time limit ($t\in [9000,10000]$). Parameters: $\mathrm{Da} = 450$, and Rankine vortex radius $a=0.05L$}
    \label{fig:fourier_vortex}
\end{figure}

{Finally, we investigated the consequences of this more complex spatiotemporal dynamics for the total abundance (proportional to the normalized abundance shown in Fig. \ref{fig:heat_map_vortex}). The diversity of possible spatiotemporal patterns makes it hard to systematically classify all the possible dynamics exhibited by the total population abundance. Instead, we discuss a few examples to illustrate the richness of behaviors that emerge in this flow with circular symmetry, which can not be aligned with the hexagonal symmetry of the natural no-flow density pattern, nor with any of its primitive vectors. At sufficiently large times, the population abundance converges to oscillations with a well-defined period. However, this often occurs after long transients, during which the population abundance changes erratically (Fig.\,\ref{fig:fourier_vortex}). In the regions of the parameter space where spots split and merge, these erratic transients in the time series of population abundance are longer, with large changes corresponding to the formation (and growth) or annihilation of spots (orange curve in Fig.\,\ref{fig:fourier_vortex}A). This erratic time series fluctuates in time without a well-defined characteristic frequency. When $\mathrm{Pe}$ increases, the central ring becomes a stable structure in which aggregates of population density do not split and merge. As a result, the population abundance oscillates in time with a characteristic oscillation frequency (larger-$\mathrm{Pe}$ curves in Fig.\,\ref{fig:fourier_vortex}), similar to what we observed for the sine flow. Finally, the normalized population abundance is always greater than 1, meaning that the inhomogeneous conditions induced by the Ranking vortex flow lead to lower population sizes as compared with the uniform solution.}

\subsection{The logistic model with nonlocal competition and facilitation}
We perform the linear stability analysis of this second model following the same steps that in Section \ref{sec:noflow}. The uniform steady solutions $\rho^*$ of Eq.\,\eqref{eq:vegetation_escaled} are the roots of a transcendental equation
\begin{equation} \label{eq:nontrivialsdfrho0}
    \mathrm{Da}^+  (1-\rho^*)\rho^* = \mathrm{Da}^- \rho^*\, e^{-\Lambda \rho^*},
\end{equation}
where $\Lambda = \overline{\chi}_f - \overline{\chi}_c$ controls the number of roots. 
The trivial, zero solution $(\rho^* = 0)$ is always an equilibrium solution.
For \(\Lambda > 0\), Eq.\,\eqref{eq:nontrivialsdfrho0} has two additional non-trivial uniform solutions: a largest one \(\rho^* =\rho_+\) and a smallest one \(\rho^* =\rho_-\).  For $\Lambda>1$ there is a subcritical range (\textit{i.e.}  $\mathrm{Da}^+ <\mathrm{Da}^-$), for which they are both positive, thus leading to a bistable regime between the zero solution and \(\rho_+\). Because of that,  we will focus on the simpler case where \(\Lambda <1\) for which Eq.\,\eqref{eq:nontrivialsdfrho0} has only one solution, \(\rho^* =\rho_+\), which is positive if \(\mathrm{Da}^+ > \mathrm{Da}^-\). We can write the positive solution in terms of the Lambert $W$ function 
\begin{equation}
   \rho_+ = 1+ \frac{1}{\Lambda}W\bigg(-\Lambda \frac{\mathrm{Da}^-}{\mathrm{Da}^+} e^{- \Lambda}\bigg).
\end{equation}
which we can evaluate numerically. Therefore, in this parameter regime, the logistic model with nonlocal competition and facilitation has one trivial zero solution for $\mathrm{Da}^- > \mathrm{Da}^+$ and a non-trivial uniform positive solution $\rho^+$ when $\mathrm{Da}^+ > \mathrm{Da}^-$. This non-trivial solution emerges from zero in a transcritical bifurcation at $\mathrm{Da}^- = \mathrm{Da}^+$, and then increases monotonically as a function of the birth-to-death ratio $\mathrm{Da}^+/\mathrm{Da}^-$, approaching the value $\rho^*\rightarrow 1$ (that is the carrying capacity value because of the scaling in \eqref{eq:units_rescale2}) for large values of this ratio.

\begin{figure}[H]
    \centering
    \includegraphics[width=0.6\textwidth]{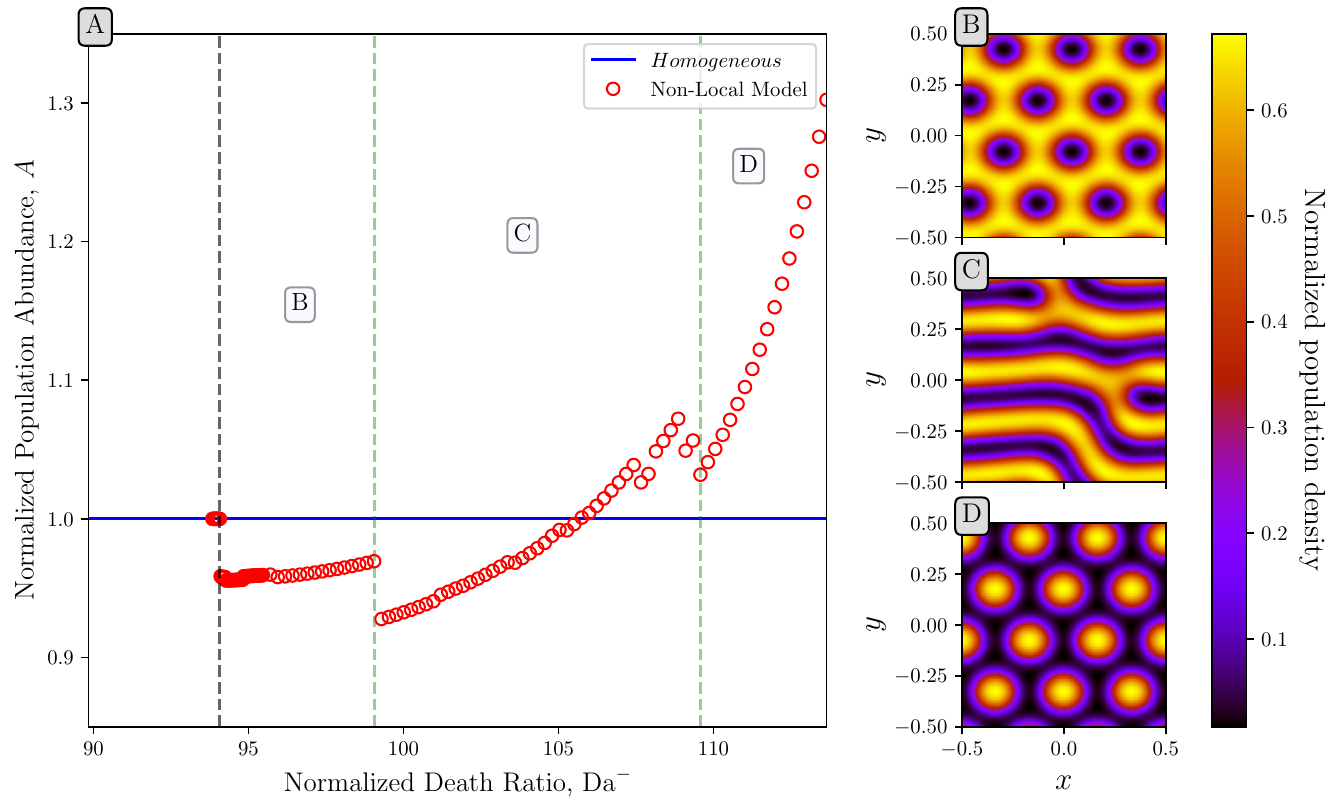}
    \caption{A) Normalized long-term population abundance as a function of $\mathrm{Da}^-$ for the patterned and uniform distributions of population density from the model in Eq.\,\eqref{eq:vegetation_escaled} in the absence of flow ($\mathrm{Pe} = 0$). Red circles are computed from simulations using $\rho_+$ plus a random perturbation as initial condition. From left to right, the black dashed line indicates the onset of pattern formation; the label (B) marks the range of $\mathrm{Da}^-$ with gap patterns; (C), labyrinths or stripes; and (D), hexagonal spots. Correspondingly, panels B, C, and D show patterns representative of each region. The unscaled parameter values: $b = 1$, $D = 10^{-4}$, $\chi_\mathrm{c}=2$, $\Lambda=0.8$, $R_\mathrm{c}=0.11$, $R_\mathrm{f}= 5 R_\mathrm{c}/14$, $K = 1$ and $L = 1$. This parameter values lead to the scaled birth rate $Da^+ \approx 121$ where the uniform population becomes zero when $\mathrm{Da}^- = \mathrm{Da}^+$}
    \label{fig:2}
\end{figure}

To find under which conditions this model might exhibit spatial patterns, we perform the linear stability analysis around the nontrivial uniform solution $\rho_+$. Considering a perturbed solution $\rho({\bf u},\tau) = \rho_+ + \epsilon \psi({\bf u},\tau)$ with $\epsilon \ll 1$, the dynamics of the perturbation in the linear regime in the absence of flow (Eq.\,\eqref{eq:vegetation_escaled} with $\mathrm{Pe}=0$) is given by:
\begin{align}\label{eq:linpertsdf}
    \frac{\partial \psi({\bf u},\tau)}{\partial \tau} = \nabla^2 \psi -\mathrm{Da}^- e^{\overline{\chi}_\mathrm{c}  \rho_+} (\psi +\rho_+\overline{\chi}_\mathrm{c} \tilde{\psi}_\mathrm{c} ) +\mathrm{Da}^+\,e^{\overline{\chi}_\mathrm{f}  \rho_+}\lbrack(1-2\rho_+)\psi +(1-\rho_+)\rho_+\overline{\chi}_\mathrm{f} \tilde{\psi}_\mathrm{f}\rbrack
\end{align}
where we have omitted the dependency of the perturbation on space and time on the right side, and $\tilde{\psi}_{\mathrm{f,c}}$ indicate the convolution of the perturbation with the corresponding kernel $G_{\mathrm{f,c}}$. Fourier transforming Eq.\,\eqref{eq:linpertsdf} and using again the ansatz $\hat{\psi}(\bm{k},\tau)\propto \exp\left(-\lambda(\bm{k})\tau\right)$ we obtain an expression for the perturbation growth rate
\begin{equation}\label{eq:pertsdf}
	\lambda(\bm{k}) = -k^2 -\mathrm{Da}^- e^{\overline{\chi}_c \rho_+}\left[1+\rho_+\overline{\chi}_c \hat{H}_c(\bm{k})\right] 
	+\mathrm{Da}^+\,e^{\overline{\chi}_f \rho_+} \left[ 1- 2\rho_++ \rho_+(1 -\rho_+)\overline{\chi}_f \hat{H}_f(\bm{k})\right],
\end{equation}
where $\hat{H}_{\mathrm{f,c}}(\bm{k})=e^{-(R_{\mathrm{f,c}} k)^2/4}$ are the Fourier transforms of the Gaussian kernels. Imposing the conditions for a Turing instability in Eq.\,\eqref{eq:pertsdf}, that is $\lambda(k_c)\geq 0$ where $k_c\neq0$ is the fastest growing mode of the perturbation, we can numerically obtain the pattern formation onset, $\mathrm{Da}^-_\mathrm{c}$ (gray dashed line in Fig.\,\ref{fig:2}), which is in excellent agreement with the numerical simulation we performed in the full nonlinear equation with $\mathrm{Pe}=0$. Note that here we focus on this transition from uniform to spatial pattern states when increasing mortality $\mathrm{Da}^-$. This is different from the only pattern-forming instability occurring in the FKPP model, in which spatial patterns appear when increasing growth, \textit{i.e.} reducing mortality. This difference probably indicates that spatial patterns in this parameter regime for this model are a signature of stress rather than a mechanism for increasing productivity. We do not consider here in detail the disappearance of patterns that will occur at larger values of $\mathrm{Da}^-$, and lead to population extinction, a behavior more similar to that of the FKPP model studied above. Our simulations show that this logistic model with nonlocal competition and facilitation, even in the absence of flow, exhibits a much higher diversity of spatial patterns than the nonlocal FKPP and more complex spatiotemporal dynamics \cite{Tlidi2008}. More specifically, as $\mathrm{Da}^-$ increases, representing worsening environmental conditions through a higher mortality rate, the model shows a sequence of gaps, labyrinths, and spots (Fig.\,\ref{fig:2}B-D), with regions of multistability between these pattern solutions \cite{Tlidi2008}. Within each pattern region, the normalized abundance increases as $\mathrm{Da}^-$ increases, but it decreases abruptly in the transitions between pattern shapes. The relative abundance can even take values below 1. Again, these behaviors contrast with the ones in the FKPP---both with flow and without it---where the abundance in the states with spatial patterns always exceeds that of the corresponding (i.e., same $\mathrm{Da}$) homogeneous states. As expected, however, the unnormalized total abundance decreases with increasing normalized death rate $\mathrm{Da}^-$.

\subsubsection{Effects of environmental flows on the pattern instability}
\label{sec:SDFresults}
\begin{figure}
    \centering
    \includegraphics[width=1\textwidth]{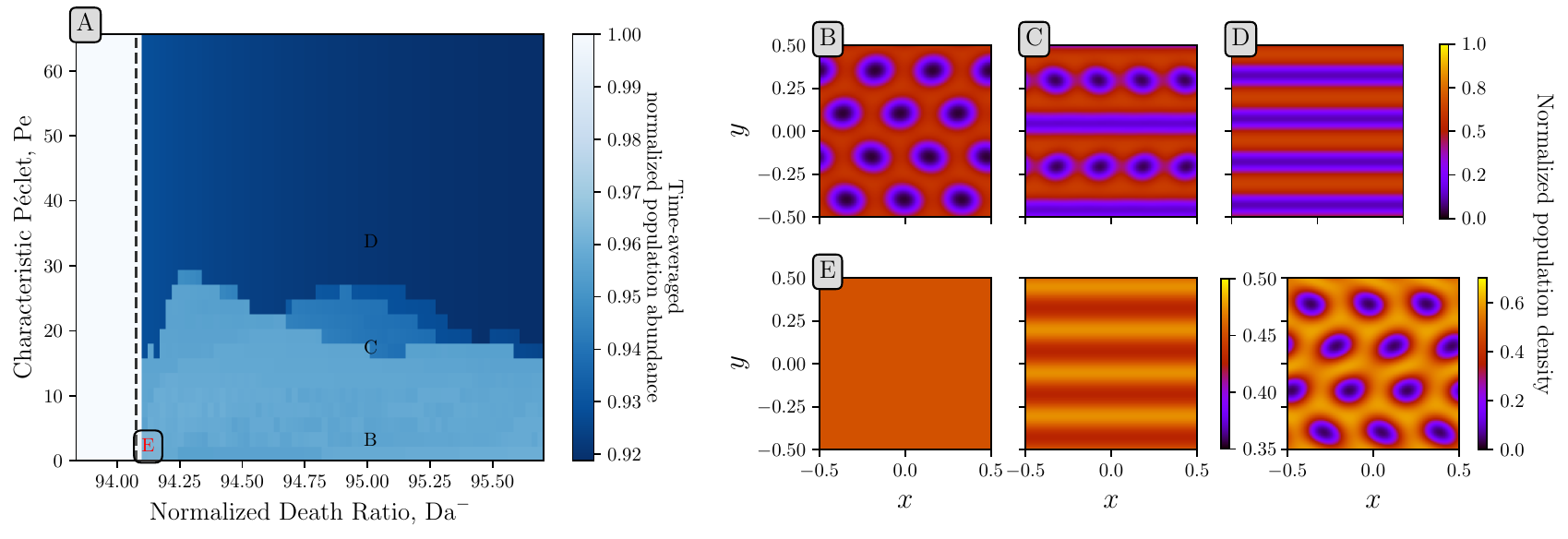}
    \caption{A) Normalized Long-term total population abundance averaged over time as a function of $\mathrm{Da}^-$ and $\mathrm{Pe}$ from simulations of model \eqref{eq:vegetation_escaled}. We performed the time average for $t>5000$, sampling the spatial configuration until the relative standard error of the time average was below $1\%$. The black dashed line indicates the pattern instability $\mathrm{Da}^-_\mathrm{c}$ predicted by the linear stability analysis, both with and without flow. Darker blue represents lower abundance relative to the uniform case (white). B-D) Snapshots of the spatial pattern once the system reaches an asymptotic state: (B) gaps moving horizontally ($\mathrm{Da}^- = 94.98, \mathrm{Pe} = 2.18$), (C) coexisting gaps and stripes ($\mathrm{Da}^- = 94.98,  \mathrm{Pe} = 16.40$), and (D) stripes ($\mathrm{Da}^- = 94.98,  \mathrm{Pe} = 32.81$). E) Pattern regimes near the onset instability at $\mathrm{Pe} = 1.64$: homogeneous distribution ($\mathrm{Da}^- = 94.05$), stripes ($\mathrm{Da}^- = 94.1$), and moving gaps ($\mathrm{Da}^- = 94.26$). Remaining parameters as in Fig.\,\ref{fig:2}.} 
    \label{fig:heat_mapSineFC}
\end{figure}

Because the linear stability analysis we performed on the nonlocal FKPP equation with $\mathrm{Pe} \neq 0$ can be generalized to the logistic model with nonlocal competition and facilitation, we expect that the sine flow will leave the pattern formation instability threshold unchanged. We confirmed this prediction with numerical simulations of Eq.\,\eqref{eq:vegetation_escaled} using the sine flow in which we varied $\mathrm{Pe}$ and kept $\mathrm{Da}^-$ within the range of values for which the model exhibits gap patterns if $\mathrm{Pe}=0$ (Fig.\,\ref{fig:heat_mapSineFC}A). We found three possible spatial configurations as $\mathrm{Pe}$ increases and the stronger shear stretches the gaps of the original pattern. When $\mathrm{Pe}$ is low, the flow is weak and it only distorts the circular shape of the gaps in the $\mathrm{Pe}=0$ pattern (Fig. \,\ref{fig:heat_mapSineFC}B). As $\mathrm{Pe}$ increases, the gaps subjected to stronger shear change to stripes until we recover a stripe pattern when $\mathrm{Pe}$ is sufficiently high (Fig. \,\ref{fig:heat_mapSineFC}C, D). Increasing the flow strength (increasing $\mathrm{Pe}$) reduces the abundance with respect to the no-flow ($\mathrm{Pe}=0$) case. We also verified, performing numerical simulations in a very small region around the pattern instability, that the first instability appearing on the homogeneous state leads to stripes, as predicted by the linear stability analysis similar to the FKPP case (Fig.\,\ref{fig:heat_mapSineFC}E-G).

\subsubsection{Rankine vortex flow}
\begin{figure}
    \centering
    \includegraphics[width=0.8\textwidth]{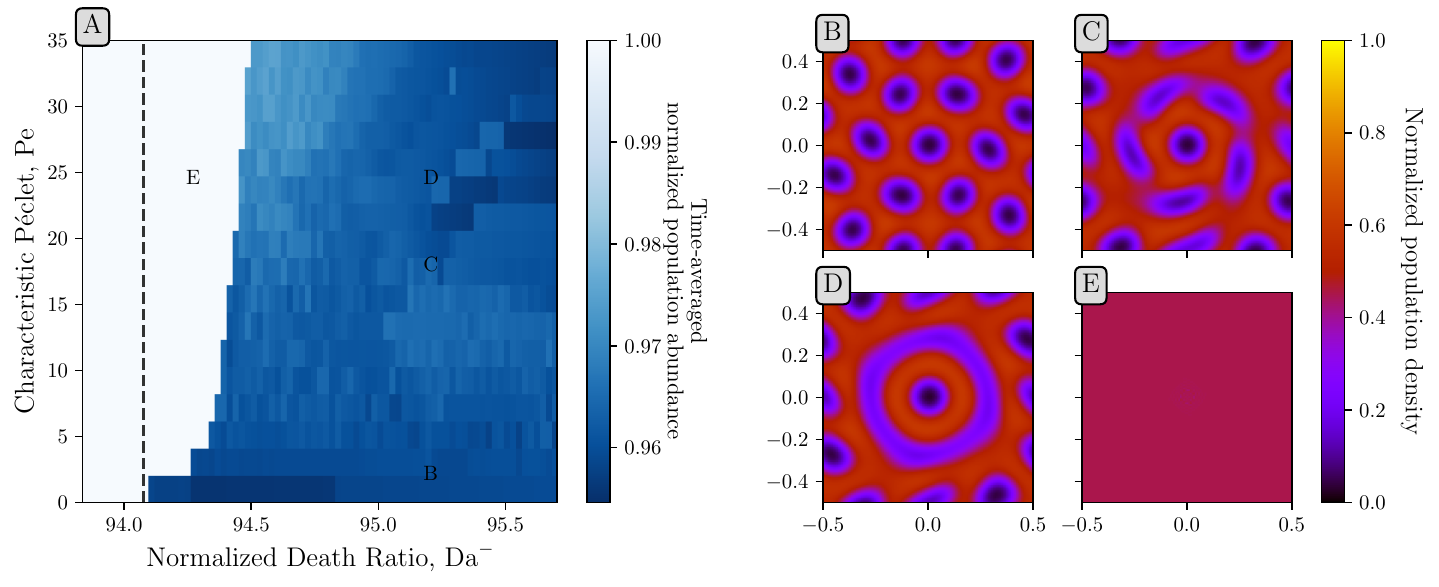}
    \caption{A) Normalized long-term population abundance averaged over time, computed as in Fig.\,\ref{fig:heat_map}, as a function of $\mathrm{Da}^-$ and $\mathrm{Pe}$ from simulations of model \eqref{eq:vegetation_escaled} with the Rankine vortex flow. The black dashed line indicates the value of the normalized death ratio, $\mathrm{Da}_c^-$, above which spatial patterns form without flow. B-D) Snapshots of typical long-term spatial pattern found in the model: (B) rotating gaps ($\mathrm{Da}^- = 95.17,\, \mathrm{Pe} = 1.64$);  (C) strecthed gaps around the center ($\mathrm{Da}^- = 95.17,\, \mathrm{Pe} = 17.5$); (D) central gap sorrounded by a ring ($\mathrm{Da}^- = 95.17,\,\mathrm{Pe} = 24.06$); and (E) Homogeneous mixing state ($\mathrm{Da}^- = 94.24, \,\mathrm{Pe} = 24.06$). Rankine vortex radius $a = 0.0312L$. Remaining parameters as in Fig.\,\ref{fig:2}. }
    \label{fig:heat_mapRankFC}
\end{figure}

In the logistic model with nonlocal competition and facilitation, the vortex flow also shifts the pattern instability toward higher values of $\mathrm{Da}^-$, but we did not find this shift to saturate within the range of $\mathrm{Pe}$ we used in the simulations (Fig.\,\ref{fig:heat_mapRankFC}A, B).
Once patterns form, the radial symmetry of the flow induces spatiotemporal patterns very similar to those in the nonlocal FKPP equation, but for gaps instead of spots. When $\mathrm{Pe}$ is low, the gaps forming the hexagonal pattern rotate around a central gap (Fig.\,\ref{fig:heat_mapRankFC}B). Then, as $\mathrm{Pe}$ increases, the higher shear stretches the gaps and arranges them circularly (Fig.\,\ref{fig:heat_mapRankFC}C). Finally, high $\mathrm{Pe}$ values destroy the gaps and create a pattern of concentric rings in the center of the environment. Outside these central rings, the population follows a more complex rotating pattern (Fig.\,\ref{fig:heat_mapRankFC}D). 

\color{black}

\section{Discussion}\label{sec:discussion}
{We investigated, using two spatially nonlocal models of population dynamics, how transport and shear in a flow environment reshape spatial patterns of population density and how these pattern changes are reflected in the dynamics of the population abundance. We found that the spatial structure of the flow velocity is a key driver of the pattern formation instability, such that flows with a spatial structure that can not be aligned with the shapes of the spatial patterns naturally arising in the population without flow, nor with their primitive vectors, enhance mixing and can stabilize homogeneous solutions that are unstable without the flow. In addition to this qualitative impact on the pattern formation instability, environmental flows reshape population patterns and impact long-term abundances.} Weak flows make patterns non-stationary by transporting and stretching density spots, which leads to different types of oscillations in the population abundance depending on the relationship between the spatial structures of the pattern and the flow. Stronger flows stretch population aggregates sufficiently to merge them with their closest neighbors, resulting in a stationary pattern in which the population abundance does not oscillate and whose shape follows the flow streamlines. 

Describing the environmental conditions and biophysical processes taking place as patterns form is thus important not only to characterize pattern formation \textit{per se}, but also to better understand its ecological consequences. Here, we considered two single-species models in which patterns form either to minimize competition \cite{Hernandez-Garcia2004} or due to the interplay between positive and negative intraspecific interactions acting at different spatial scales \cite{Tlidi2008}. 
Flows perturb the equilibrium configurations by making population spots interact with each other, which makes population abundances tend to its homogeneous value as flow intensity increases. In cases with more complex ecological dynamics, including communities with different interactions, flow-induced changes in the spatial patterns of population density may shift the equilibrium state in different directions. Previous work has shown that spatial patterns induced by nonlocal interactions can prevent competitive exclusion by segregating species in space \citep{Maciel2021, Simoy2023} or change the demographic properties of Allee effects \citep{Jorge2024}. Our results suggest that these pattern-induced phenomena could be very sensitive to the presence of flows. Flows could, for example, prevent competing species from segregating in space, as shown in \citep{Maciel2021, Simoy2023}, forcing them to interact periodically and ultimately causing the extinction of the weaker competitor. {The importance of the flow spatial structure in determining the pattern formation instability is ecologically relevant too. In scenarios where spatial patterning is necessary to stabilize ecological communities, flow environments tending to stabilize homogeneous distributions of organisms could lead to species extinctions, whereas other flow structures could favor diversity.}

Besides choosing a simple ecological scenario, we simplified the mixing process by choosing incompressible, stationary, {and spatially periodic flows. For example, the sine flow approximates some of the features of laminar, parabolic flows in confined channels, but both flows differ in some important dynamical features due to the role of viscosity and boundaries. Similarly, different boundary conditions can also impact patterning on the different ecological models \cite{Dillon1994}.} Non-stationary flows enable a much richer variety of mixing processes, including chaotic regimes whose statistical properties mimic those of turbulent environments \citep{Ottino1990}. In these regimes, the flow can create regions of different mixing, including ones where organisms remain trapped for a longer time and interact more often with their neighbors \citep{Zaslavsky2002,Neufeld2009}. This flow-induced heterogeneity in species interactions can impact ecological processes, such as plankton blooms \citep{Reigada2003} and pathogen transmission \citep{Brookes2004}, and alter the outcome of evolutionary dynamics \citep{Karolyi2005, Miranda2023}. Transport properties also change when considering compressible flows, which impacts population mixing \citep{Vergassola1997,Volk2014}. Previous work in local models of population dynamics has shown that considering mixing by compressible flows can importantly impact both the spatial distribution of organisms and the long-term ecological and evolutionary dynamics \citep{Perlekar2010,Pigolotti2012,Benzi2012,Plummer2019}. Therefore, investigating the emergent spatial patterns and population dynamics due to nonlocal interactions in environments with compressible or chaotic flows is an interesting direction for future research.

\section*{Acknowledgments}
This work was partially funded by the Center of Advanced Systems Understanding (CASUS), which is financed by Germany’s Federal Ministry of Education and Research (BMBF) and by the Saxon Ministry for Science, Culture and Tourism (SMWK) with tax funds on the basis of the budget approved by the Saxon State Parliament. N.O.S was partially supported by Coordenação de Aperfeiçoamento de Pessoal de Nível Superior (CAPES) - Finance Code 001 and by the CAPES-Print program through a sandwich doctoral fellowship at CASUS, Germany. RMG and JV were partially supported by FAPESP through a BIOTA Jovem Pesquisador Grant 2019/05523-8 (RMG and JV), ICTP-SAIFR grant 2021/14335-0 (RMG), and a Master's fellowship 2020/14169-0 (JV). RMG acknowledges support from Instituto Serrapilheira (Serra-1911-31200). C.L. and E.H-G were supported by grants LAMARCA PID2021-123352OB-C32 funded by MCIN/AEI/10.13039/501100011033323 and FEDER "Una manera de hacer Europa"; and TED2021-131836B-I00 funded by MICIU/AEI/\,\newline
10.13039/501100011033 and by the European Union "NextGenerationEU/PRTR". E.H-G. and N.O.S. also acknowledge the Maria de Maeztu program for Units of Excellence, CEX2021-001164-M funded by MCIN/AEI/10.13039/501100011033. C.L. was partially supported by the Scultetus Center Visiting Scientist Program at CASUS.

\bibliography{sn-bibliography_doi}

\newpage 

\setcounter{figure}{0}      
\renewcommand\thefigure{A\arabic{figure}}    
\section*{Appendices}
\begin{appendices}

\section{Pseudospectral method to integrate Eq.\,\eqref{eq:Fkkp_advection}.}\label{app:numerical}
We numerically integrate Eqs.\,\eqref{eq:Fkkp_advection} and \eqref{eq:veg_model} using a pseudospectral integration method on a spatial grid with periodic boundary conditions, cell size $\Delta x$ and time step $\Delta t$. The key idea of the pseudospectral approach is to decompose the partial differential equation (PDE) into its linear and nonlinear components,
\begin{equation}
    \label{app:Fkkp_advection}
    \partial_t \rho({\bf x}, t) = \mathcal{L} \rho({\bf x}, t) + \mathcal{N}[\rho({\bf x}, t)],
\end{equation}
where $\mathcal{L}$ is a linear differential operator (e.g., diffusion, advection, linear growth), and $\mathcal{N}(\rho({\bf x}, t))$ encapsulates nonlinear terms (e.g., ecological interactions, nonlinear transport).

Next, we apply the fast Fourier transform (FFT) to Eq.\,\eqref{app:Fkkp_advection}, converting it into an ordinary differential equation in Fourier space:
\begin{equation}
    \label{app:Fkkp_advection_Fourier}
    \partial_t \hat{\rho}({\bf k}, t) = \alpha({\bf k}) \hat{\rho}({\bf k}, t) + \Phi({\bf k}, t)
\end{equation}
where $\hat{\rho}({\bf k}, t) = \mathcal{F}[\rho({\bf x}, t)]$ is the Fourier transform of $\rho({\bf x}, t)$; $\alpha({\bf k})$ is the Fourier linear coefficient of $\mathcal{L}$; and $\Phi({\bf k}, t)$ is the Fourier transform of the nonlinear term, $\mathcal{N}[\rho({\bf x}, t]$. We integrate the resulting ODE in Fourier space Eq.\,\eqref{app:Fkkp_advection_Fourier} using a fourth-order Runge-Kutta (RK4) scheme:
\begin{enumerate}
    \item {Nonlinear evaluation}: at each substep, \(\rho({\bf x}, t)\) is reconstructed via inverse FFT, \(\mathcal{N}(\rho)\) is computed in real space, and then transformed back to Fourier space.
    \item {Linear evolution}: the linear term \(\alpha({\bf k}) \hat{\rho}({\bf k}, t)\) is handled exactly in Fourier space.
    \item {Time stepping}: the RK4 scheme combines intermediate steps to advance \(\hat{\rho}({\bf k}, t)\) with \(\mathcal{O}(\Delta t^4)\) accuracy.
\end{enumerate}
The Python code with the algorithm implementation is available at \href{https://github.com/Jaegg3rNat/NonLocalShear.git}{GitHub}.

\section{Linear stability analysis of the nonlocal logistic model with a sine flow}\label{app:LSA-flow}

 To perform the linear stability analysis of the nonlocal logistic model with a sine flow [Eqs.\,\eqref{eq:vx_sin}-\eqref{eq:vy_sin} with $m=1$], we first linearize Eq.\,\eqref{eq:Fkkp_advection} around the uniform configuration assuming a perturbed solution $\rho({\bf u},\tau) = \rho^* + \epsilon\psi({\bf u},\tau)$ where $ \rho^* = \mathrm{Da}$ is the uniform solution, and $\epsilon\ll 1$. This linearization gives an equation for the perturbation,
\begin{equation}\label{eq:app-linear-flow}
      \frac{\partial \psi({\bf u},\tau)}{\partial \tau} =  \boldsymbol{\nabla}^2_u  \psi({\bf u},\tau) - \,\mathrm{Da} \int_E H({\bf u-v})\psi({\bf v},\tau) \,\mathrm{d}{\bf v}- \, \mathrm{Pe}\, \sin\left(\tilde{q}\, v\right)\frac{\partial \psi({\bf u},\tau)}{\partial u}.
\end{equation}
 
Next, because the system is periodic, we write an ansatz for the perturbation using a Floquet decomposition \citep{Deconinck2006,jordan2007}
\begin{equation}\label{eq:floquet}
    \psi({\bf u},\tau)\equiv \psi_{{\bf k}}({\bf u},\tau) = {\rm e}^{i{ {\bf k \cdot {\bf u}}+\lambda({\bf k})\tau}}\sum_n C_n({\bf k}){\rm e}^{i n \tilde{q} v}.
\end{equation}
Because we are using $m=1$, the sine flow has the same periodicity as the system domain. and the Floquet expansion in Eq.\,\eqref{eq:floquet} can be rearranged to become a simple Fourier expansion. Here, we keep this slightly more general formulation because it could be generalized without difficulty to $m>1$. Inserting the ansatz in Eq.\eqref{eq:floquet} into the linear equation for the perturbation, Eq.\,\eqref{eq:app-linear-flow}, we obtain the following four terms.
\begin{itemize}

    \item[-] For the time derivative, we get
        \begin{equation}
             \partial_\tau \psi = e^{i {\bf k}\cdot {\bf u} + \lambda({\bf k}) \tau}\; \sum_n \lambda({\bf k}) C_n ({\bf k}) e^{in \tilde{q} v}
        \end{equation}
        
    \item[-] For the diffusion term,
        \begin{equation}
            \boldsymbol{\nabla}^2_u  \psi({\bf u},\tau)  = - e^{i {\bf k}\cdot {\bf u} + \lambda({\bf k}) \tau} \sum_n k^2_n  C_n ({\bf k}) e^{in \tilde{q} v},
        \end{equation}
        where we have defined ${\bf k}_n=(k_u,k_v+n\tilde{q})$ and $k_n=\rvert {\bf k}_n \rvert$.
        
    \item[-] For the nonlocal competition term, the convolution becomes
        \begin{equation}
            \int_E H({\bf u-v})\psi({\bf v},\tau) \,\mathrm{d}{\bf v} =  e^{i {\bf k}\cdot {\bf u} + \lambda({\bf k}) \tau} \sum_n \hat{H}(k_n)\,C_n ({\bf k}) e^{in \tilde{q} v}.
        \end{equation}
        
    \item[-] Finally, we can compute the advection term by expanding the sine function in the Euler notation and renaming the index $n$ in the series This procedure leads to
        \begin{equation}
            \mathrm{Pe}\, \sin\left(\tilde{q}\, v\right)\frac{\partial \psi({\bf u},\tau)}{\partial u} = \frac{\mathrm{Pe}}{2} k_u  e^{i {\bf k}\cdot {\bf u} + \lambda({\bf k}) \tau} \sum_n \Big[ C_{n+1}({\bf k}) - C_{n-1}({\bf k})\Big] e^{in \tilde{q} v}
        \end{equation}
        
\end{itemize}

\noindent Collecting all these terms, we obtain an infinite-dimensional system of coupled equations for the coefficients $C_n ({\bf k})$
\begin{equation}\label{eq:appinf-sys}
    \left[\lambda({\bf k}) + k_n^2 + \mathrm{Da} \, \hat{H}(k_n)\right] C_n({\bf k}) = - k_u \frac{\mathrm{Pe}}{2} \Big[C_{n+1}({\bf k})-C_{n-1}({\bf k})\Big].
\end{equation}
 Notice that, for $\mathrm{Pe}=0$, this system of equations (\ref{eq:appinf-sys}) reduces for $n=0$ to the expression we obtained for $\lambda(k)$ in the absence of flow,
\begin{equation}
\label{eq:b8}
    \lambda(k) = -k^2 - \mathrm{Da}\, \hat{H}(k)\,.
\end{equation}
{Evaluating this limit case, we can obtain $\mathrm{Da}_\mathrm{c}$ by computing the value of this parameter at which the maximum of $\lambda(k)$ becomes positive (cyan dashed line in Fig.\,\ref{fig:eigen}). We obtain the same expression \eqref{eq:b8} if $k_u=0$, which corresponds to perturbations with the shape of horizontal stripes. This implies that, as we increase $\mathrm{Da}$, a stripe pattern-forming instability emerges when $\mathrm{Da}=\mathrm{Da}_\mathrm{c}$, the critical Damköhler value in the absence of flow. 

To explore whether additional instabilities arise at lower $\mathrm{Da}$, we can further analyze Eq.\,(\ref{eq:appinf-sys}) using a low-$\mathrm{Pe}$ perturbation theory. Because the $n=0$ eigenvalue branch is the relevant one, we can perform such small-$\mathrm{Pe}$ perturbation theory around this $n=0$ branch, which involves the Fourier coefficient $C_0({\bf k})$. From Eq.\,\eqref{eq:appinf-sys}, only the coefficients $C_1({\bf k})$ and $C_{-1}({\bf k})$ are coupled directly (i.e. to order $\mathrm{Pe}$) to $C_0({\bf k})$. Other coefficients with $\rvert n \rvert \geq 2$ are only coupled indirectly to $C_0({\bf k})$ and thus order $\mathrm{Pe}^2$ or higher. Following this reasoning, we can truncate the sum in $n$ at $n =\pm 1$, which leads to a reduced system that we write in matrix form as} 
\begin{equation}\label{eq:sys-matrix}
   \begin{pmatrix}
        \lambda({\bf k}) +A(k_{-1}) &-B & 0 \\
B& \lambda({\bf k}) +A(k_0)&-B\\
 0& B&\lambda({\bf k}) + A(k_1)\\       
    \end{pmatrix}
    \begin{pmatrix}
    C_{-1}({\bf k})\\
        C_{0}({\bf k})\\
        C_{1}({\bf k})\\
     \end{pmatrix} = 0
\end{equation}
with $A(k_n)=k_n^2 + \nu \hat{H}(k_n)$ and $B=k_u \mathrm{Pe}/2$. The eigenvalues of the matrix of coefficients in Eq.\,\eqref{eq:sys-matrix} can be computed analytically from its determinant. Its expressions, however, involve the sum of several Bessel functions with shifted arguments. We obtained them using Mathematica \citep{Mathematica} and computed the $\mathrm{Da}$ value at which the maximum of the real part of the largest eigenvalue becomes positive. As shown in Fig.\,\ref{fig:eigen}, this critical value of $\mathrm{Da}$ is the same we obtained in the absence of flow, which indicates that the sine flow has no effect on determining the pattern instability. This is consistent with the results of the nonlinear numerical simulations, which confirm that the homogeneous solution remains stable for $\mathrm{Da}<\mathrm{Da}_\mathrm{c}$.

\section{Supporting Figures}
\renewcommand\thefigure{C\arabic{figure}}    
\setcounter{figure}{0}      

\begin{figure}[H]%
\centering
\includegraphics[width=0.9\textwidth]{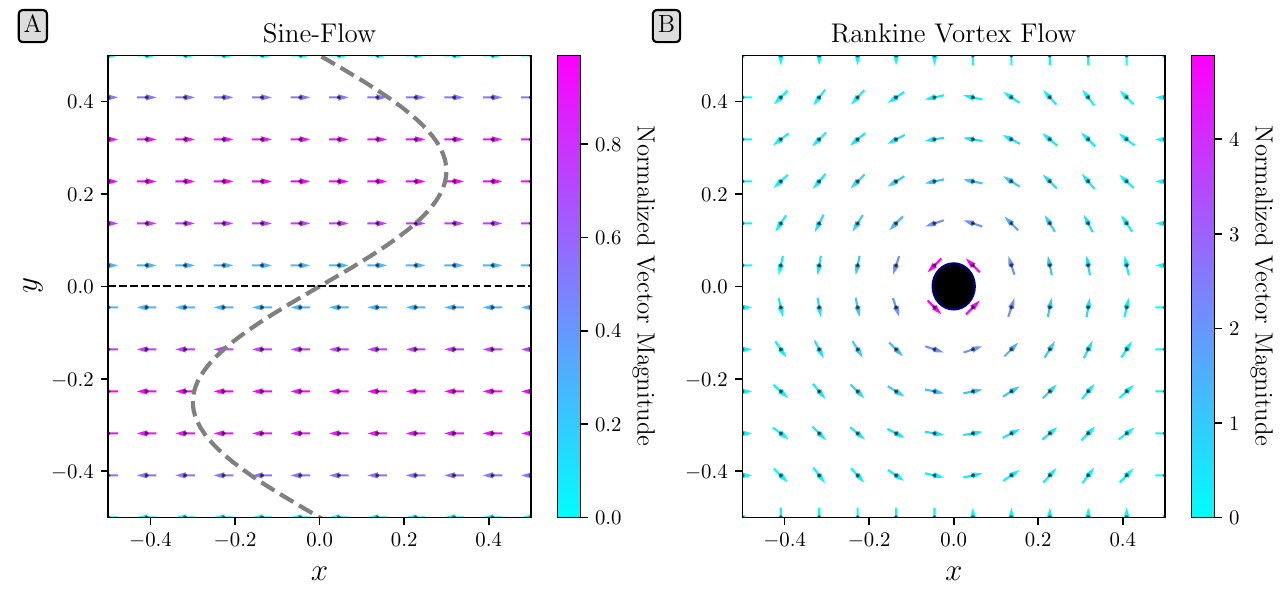}
\caption{Vector field of the (A) sine and (B) Rankine vortex flow. The arrows indicate the direction of the flow, and the intensity is given by the color as indicated in each panel's colorbar. The central black disk in B marks the region where velocity increases linearly with distance to the vortex center (placed at the center of the simulation domain).}\label{fig:flowvector}
\end{figure}
\begin{figure}[H]%
\centering
\includegraphics[width=0.6\textwidth]{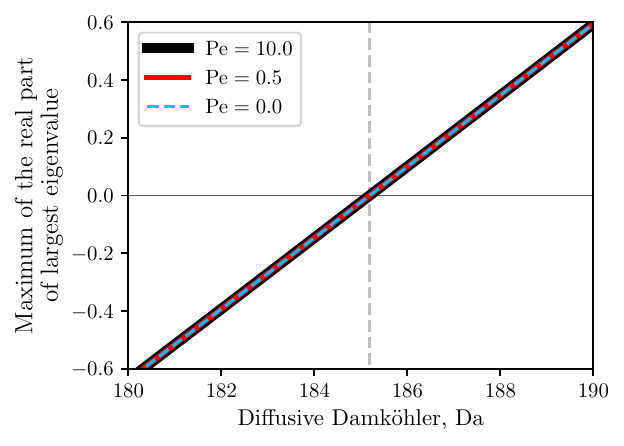}
\caption{Maximum of the real part of the largest eigenvalue as a function of $\mathrm{Da}$ for the non-flow system and two different values of $\mathrm{Pe}$. The gray-dashed vertical line indicates the value of $\mathrm{Da}_\mathrm{c}$.}\label{fig:eigen}
\end{figure}
\begin{figure}[H]%
\centering
\includegraphics[width=0.7\textwidth]{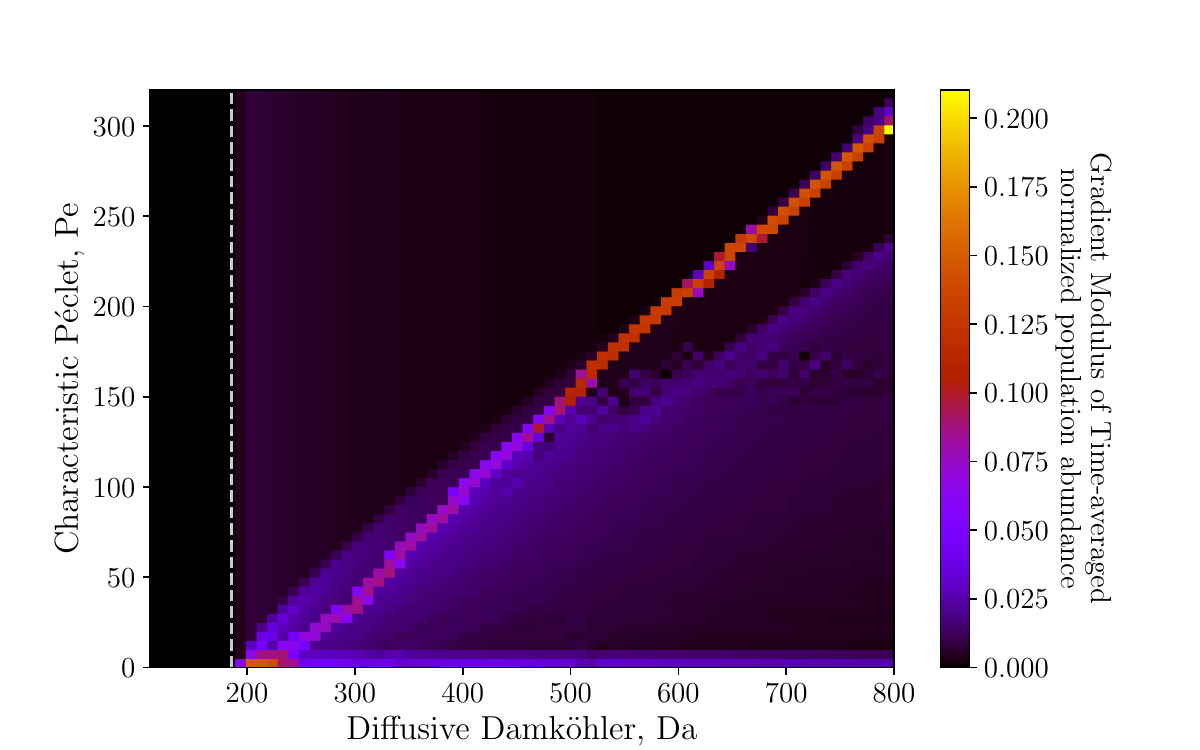}
\caption{FKPP model under the sine flow. The modulus of the gradient of the time-averaged normalized population abundance from Fig. 2 in the Pe-Da plane has sharp maxima that locate the transitions between pattern shapes and allows us to define sub-regions in the parameter space with qualitatively similar spatiotemporal population dynamics. The white dashed line indicates the value of the Damköhler number $\mathrm{Da}_\mathrm{c}$ at which patterns first form in the absence of flow.}\label{fig:SM1}
\end{figure}
\begin{figure}[H]%
\centering
\includegraphics[width=0.7\textwidth]{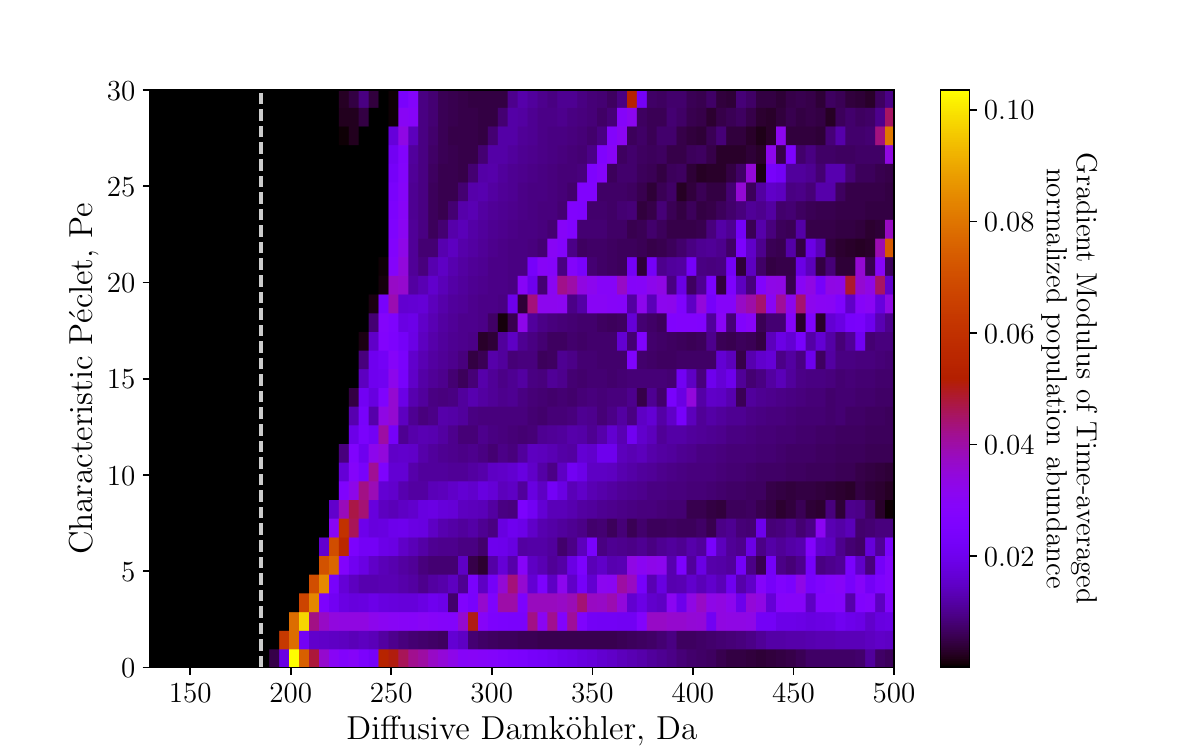}
\caption{FKPP model under the Rankine vortex. The modulus of the gradient of the time-averaged normalized population abundance from Fig. 4 in the Pe-Da plane does not display a structure clear enough to capture the transitions between pattern shapes and hence does not allow us to define sub-regions in the parameter space with qualitatively similar spatiotemporal population dynamics. The white dashed line indicates the value of the Damköhler number $\mathrm{Da}_\mathrm{c}$ at which patterns first form in the absence of flow.}\label{fig:SM2}
\end{figure}


\end{appendices}

\end{document}